%% file: paper.tex
\documentclass[11pt]{article}

\usepackage[english]{babel}
\usepackage[a4paper,dvips,nohead,twoside]{geometry}
\usepackage{ucs}
\usepackage[utf8x]{inputenc}
\usepackage{amsmath,amssymb}
\usepackage[dvips]{graphicx}
\usepackage{psfrag}
\usepackage{xspace}
\usepackage{tensor}
\usepackage[title]{appendix}

\usepackage[longversion]{optional}

\usepackage{xr}
\usepackage[ps2pdf, bookmarks=true,
            bookmarksopen=true,colorlinks=false]{hyperref}

\hypersetup{%
  pdftitle = {Non-genericity of the Nariai solutions:
    II. Investigations within the Gowdy class}, 
  pdfsubject = {},
  pdfauthor = {Florian Beyer}, 
  pdfkeywords = {}%
}

 \usepackage[thmmarks,amsmath,hyperref]{ntheorem}

 \theoremstyle{plain}
 \theorembodyfont{\normalfont}
 \theoremsymbol{\ensuremath{_\Box}}

 \theoremheaderfont{\sc}\theorembodyfont{\small\upshape}
 \theoremstyle{nonumberplain}
 \theoremseparator{:}
 \theoremsymbol{\rule{1ex}{1ex}}
 \theoremstyle{break}
 \theoremheaderfont{\normalfont\small}
 \theorembodyfont{\it\small\flushleft}
 \theoremsymbol{}
 
 \theorembodyfont{\it\tiny\flushleft}

\newcommand{\R}{\ensuremath{\mathbb R}\xspace}

\newcommand{\U}{\ensuremath{\text{U}(1)}\xspace}
\newcommand{\T}{\ensuremath{\mathbb{T}^3}\xspace}
\newcommand{\SO}{\ensuremath{\text{SO}(3)}\xspace}
\newcommand{\So}{\ensuremath{\mathbb S^1}\xspace}
\newcommand{\St}{\ensuremath{\mathbb S^2}\xspace}
\newcommand{\Sth}{\ensuremath{\mathbb S^3}\xspace}
\newcommand{\SoXSt}{\ensuremath{\mathbb S^1\times\mathbb S^2}\xspace}

\newcommand{\scrip}{\ensuremath{\mathcal J^+}\xspace}
\newcommand{\scrim}{\ensuremath{\mathcal J^-}\xspace}

\newcommand{\normconstr}{\mathrm{Norm}^{(\textit{constr})}}
\newcommand{\normeinstein}{\mathrm{Norm}^{(\textit{einstein})}}
\newcommand{\normbc}{\mathrm{Norm}^{(\textit{BC})}}

\newcommand{\Connection}[3]{\Gamma\indices{_{#1}^{#2}_{#3}}}
\newcommand{\Eqref}[1]{Eq.~\eqref{#1}}
\newcommand{\Eqsref}[1]{Eqs.~\eqref{#1}}
\newcommand{\Sectionref}[1]{Section~\ref{#1}}

\newcommand{\Figref}[1]{Fig.~\ref{#1}}

\graphicspath{{figures/}}
\numberwithin{equation}{section}

\begin{document}
\title{%
  Non-genericity of the Nariai solutions:\\
  II.\ Investigations within the Gowdy class
}

\author{Florian Beyer\\\textit{\small beyer@ann.jussieu.fr}}
  
\date{{\small Laboratoire Jacques-Louis Lions\\
Universit\'e Pierre et Marie Curie (Paris 6)\\
4 Place Jussieu, 75252 Paris, France}}

\maketitle

\begin{abstract}
\input{abstract}
\end{abstract}

\renewcommand{\figurename}{Fig.}
\bibliographystyle{hplain}

\section{Introduction}
\input{introduction}

\section{Preparations}
\label{sec:preparations}

\input{gowdysymmetry}

\input{initial_data}

\input{expansions}

\input{formulation_numerics}

\section{Results}
\label{sec:numresults}
\input{numresults_smallamplitudes}

\section{Summary and outlook}
\label{sec:summary}
\input{summary}

\section{Acknowledgments}
\input{acknowledgements}

\opt{longversion}{%
  \begin{appendices}
    \input{gowdysymmetry_appendix} 
  \end{appendices}%
}

\bibliography{bibliography}
\end{document}

%% file: abstract.tex
This is the second of two papers where we study the asymptotics of the
generalized Nariai solutions and its relation to the cosmic no-hair
conjecture.  In the first paper, the author suggested that according
to the cosmic no-hair conjecture, the Nariai solutions are non-generic
among general solutions of Einstein's field equations in vacuum with a
positive cosmological constant. We checked that this is true within
the class of spatially homogeneous solutions. In this paper now, we
continue these investigations within the spatially inhomogeneous Gowdy
case. On the one hand, we are motivated to understand the fundamental
question of cosmic no-hair and its dynamical realization in more
general classes than the spatially homogeneous case. On the other
hand, the results of the first paper suggest that the instability of
the Nariai solutions can be exploited to construct and analyze
physically interesting cosmological black hole solutions in the Gowdy
class, consistent with certain claims by Bousso in the spherically
symmetric case. However, in contrast to that, we find that it is not
possible to construct cosmological black hole solutions by means of
small Gowdy symmetric perturbations of the Nariai solutions and that
the dynamics shows a certain new critical behavior.  For our
investigations, we use the numerical techniques based on spectral
methods which we introduced in a previous publication.


%% file: introduction.tex
In this work, we are interested in a particular consequence of the
cosmic no-hair conjecture \cite{gibbons77,Hawking82}. As discussed in
our first paper \cite{beyer09:Nariai1}, this conjecture suggests that
the so-called Nariai solutions are non-generic in the class of
cosmological solutions of Einstein's field equations in vacuum
\begin{equation}
  \label{eq:EFE}
  G_{\mu\nu}+\Lambda g_{\mu\nu}=0,
\end{equation}
with a positive cosmological constant $\Lambda$, due to its
extraordinary asymptotics for large times. Throughout the paper, a
cosmological solution means a globally hyperbolic solution of
\Eqref{eq:EFE} with compact Cauchy surfaces.

In the first paper \cite{beyer09:Nariai1}, we analyzed the asymptotics
of what we called generalized Nariai solutions. A particular solution
in this family is the (standard) Nariai solution
\cite{Nariai50,Nariai51}. Although these solutions are all isometric
locally to the standard Nariai solution, we were motivated to
introduce this family for the following reasons. First, their
spatially homogeneous perturbations have interesting properties
\cite{beyer09:Nariai1}.  Second, the approach presented in this paper
here makes particular use of the non-standard Nariai solutions in this
family. In the following, we often speak of ``Nariai solutions'', when
we mean any generalized Nariai solution.  A particular contribution of
the first paper was a proof of the outstanding fact that the Nariai
solutions do not possess smooth conformal boundaries, a result closely
related to the cosmic no-hair picture as explained there.  Moreover,
we investigated how the expected non-genericity of the Nariai
solutions is realized dynamically in the spatially homogeneous class
of perturbations. In general, when we speak of a perturbation of a
Nariai solutions, we mean a cosmological solution of the fully
non-linear Einstein's field equations \Eqref{eq:EFE} whose data, on
some Cauchy surface, is close to the data on a Cauchy surface of a
generalized Nariai solution. By ``close'' we mean that two data sets
should deviate not too much with respect to some reasonable norm in
the initial data space. We show in \cite{beyer09:Nariai1} that an
arbitrary small spatially homogeneous perturbation of any Nariai
solution does either not expand at all in, say, the future, or it
expands in a manner consistent with the cosmic no-hair picture by
forming a smooth future conformal boundary.

Certainly, the case of spatially homogeneous perturbations is special
and it would be interesting to study the instability of the Nariai
solutions within more general classes of perturbations. Beyond the
problem of cosmic no-hair, however, it is a tempting possibility to
exploit our knowledge about the instability in the homogeneous case in
order to construct new non-trivial inhomogeneous cosmological black
hole solutions. Recall from the results in the first paper
\cite{beyer09:Nariai1} that in the spatially homogeneous case, the
sign of the initial expansion $H_*^{(0)}$ of the spatial \St-factor of
a perturbation of a Nariai solution controls whether the spatial
\St-factor collapses or expands to the future. Hence, one can expect
that by making $H_*^{(0)}$ spatially dependent on the initial
hypersurface, we become able to control the \textit{spatially local}
behavior of the perturbations. In particular, we should obtain
solutions with arbitrary many black hole interiors on the one hand and
expanding cosmological regions on the other hand. In principle, we are
interested in studying generic inhomogeneous perturbations of the
Nariai solutions without any symmetries. However, this is not feasible
in practice. A systematic approach would be to reduce the symmetry
assumptions step by step. The first systematic step is the spherically
symmetric case. Indeed Bousso in \cite{Bousso03} claims that
arbitrarily complicated spherically symmetric cosmological black hole
solutions can be constructed with this approach.

In this paper, we do not consider the spherically symmetric case, but
rather proceed with the Gowdy symmetric case. Gowdy symmetric
solutions with spatial \T-topology have been very prominent in the
field of mathematical cosmology, and recently, important outstanding
problems have been tackled rigorously; see the important work in
\cite{Ringstrom06,Ringstrom06b} based on a long list of previous
references. However, the Gowdy case with spatial \SoXSt-topology (and
\Sth), which is the relevant case here, has turned out to be more
difficult analytically \cite{Isenberg89,garfinkle1999,Stahl02}.  This
is one particular motivation for us to proceed with this class by
means of numerical techniques.  In fact, we have developed numerical
techniques in \cite{beyer08:code} applicable to the \SoXSt-Gowdy
class, which will be used in this work here. An alternative numerical
approach for Gowdy solutions with \SoXSt-topology can be found in
\cite{garfinkle1999}. However, this approach is not applicable
directly for the conformal field equations based on orthonormal frames
which we have decided to work with.  In any case, our results can be
hoped to complement the claims in \cite{Bousso03} in a physically and
technically interesting setting.

In all what follows, we use the same fundamental conventions and
assumptions as in the first paper \cite{beyer09:Nariai1}.

Our paper is organized as follows. In \Sectionref{sec:preparations},
we prepare our investigations. After a short introduction to Gowdy
symmetry on \SoXSt in \Sectionref{sec:Gowdy symmetry}, we construct
those Gowdy invariant initial data sets
in \Sectionref{sec:initialdata}, which will be used to study the
perturbations of the Nariai solutions later. For the basic properties
of these solutions, we refer to \Sectionref{P1-sec:cosmicnohairnariai}
of \cite{beyer09:Nariai1}. Then,
\Sectionref{sec:expansions} of the paper here is devoted to the
discussion of certain mean curvature quantities which are analogous to
those quantities which control the instability of the Nariai solution
in the spatially homogeneous case.  In \Sectionref{sec:formulation},
we present the formulation of Einstein's field equations that we will
use for the numerical computations, namely the conformal field
equations. We briefly describe the unknown variables in this
formulation and fix our choice of gauge. We conclude with a comment
about our particular reduction to $1+1$. We proceed by giving a quick
summary of the numerical infrastructure
in \Sectionref{sec:numerics}. In \Sectionref{sec:numcompID}, we
comment on how to compute initial data for the conformal field
equations from the data constructed
in \Sectionref{sec:initialdata}. Numerically, this is not completely
trivial, because the initial data before are based on coordinate
components of the metric and hence they run the risk of coordinate
singularities when they are transformed to orthonormal frame based
data for the conformal field
equations. In \Sectionref{sec:choicedata}, we fix and motivate
particular initial data sets for the later numerical runs. The central
part of the paper is \Sectionref{sec:thenumresults}, where we show the
numerical evolutions and interpret the
results. In \Sectionref{sec:DetailsErrors}, we present further details
of a practical nature about the numerical runs and then proceed with
an analysis of numerical errors. The paper is concluded with a
summary, a discussion of open problems and an outlook
in \Sectionref{sec:summary}.


%% file: gowdysymmetry.tex
\subsection{Gowdy symmetry on
  \texorpdfstring{\SoXSt}{S1xS2}}  
\label{sec:Gowdy symmetry}  
We quickly introduce important and relevant facts about Gowdy symmetry
on \SoXSt.  General aspects of $\U\times\U$-symmetric solutions of
Einstein's field equations were discussed in \cite{Gowdy73} for the
first time and later reconsidered in \cite{chrusciel1990}.

\paragraph{Smooth $\U\times\U$-invariant metrics on \SoXSt}
Let us introduce coordinates $(\rho,\theta,\phi)$ on \SoXSt, where
$\rho\in(0,2\pi)$ is the standard parameter on $\So$ and
$(\theta,\phi)$ are standard polar coordinates on \St. The coordinate
vector fields $\partial_\rho$ and $\partial_\phi$ generate a smooth
effective action of the group $\U\times\U$ on \SoXSt.  Let us consider
a smooth Riemannian metric $h$ on \SoXSt which is invariant under this
action. One can parametrize $h$ as
\begin{equation}
  \label{eq:Gowdyh}
  h=e^{2\lambda} d\theta^2+R(e^P d\phi^2+2 e^P Q d\phi\,d\rho
  +(e^{P}Q^2+e^{-P})d\rho^2).
\end{equation}
In particular, the field $\partial_\theta$ can be assumed to be
orthogonal to the group orbits everywhere. All functions involved here
only depend on $\theta$. For the smoothness of $h$, it is sufficient
and necessary \cite{chrusciel1990} that there are function $\tilde R$,
$\tilde P$ and $\tilde \lambda$ so that
\begin{subequations}
  \label{eq:GowdyS1xS2behavior}
  \begin{equation}
    R=\tilde R\sin\theta,\quad
    P=\tilde P+\ln\sin\theta,\quad
    \lambda=(\tilde P+\ln\tilde R)/2+\tilde\lambda,
  \end{equation}
  where $\tilde R$, $\tilde P$, $\tilde \lambda$ and $Q$ are smooth
  function of $\cos\theta$, and
  \begin{equation}
    \tilde\lambda\bigr|_{\theta=0,\pi}=0.
  \end{equation}
\end{subequations}

\paragraph{Gowdy symmetric spacetimes with spatial \SoXSt-topology}
Now let us consider a globally hyperbolic spacetime $(M,g)$ foliated
with $\U\times\U$-invariant Cauchy surfaces of topology \SoXSt, in the
sense that the first and second fundamental form of each surface are
invariant under the $\U\times\U$-action before.  It has been shown
before \cite{chrusciel1990,Clausen07} that this action is orthogonally
transitive, i.e. the twist constants
\[c^1 :=\epsilon^{\mu\nu\sigma\lambda}\eta_{\mu}^1\sigma_\nu^2
  \partial_\sigma\eta_{\lambda}^a,\quad
c^2 :=\epsilon^{\mu\nu\sigma\lambda}\eta_{\mu}^1\sigma_\nu^2
\partial_\sigma\sigma_{\lambda}^a
\] 
vanish for spatial topology \SoXSt (or \Sth, but not $\T$) if this
spacetime is a solution of \Eqref{eq:EFE}. Here all index
manipulations are done with the metric $g$, and $\epsilon$ is the
volume form of $g$ assuming that the spacetime is orientable. See also
\cite{Wainwright} for more details. In general, if these constants
vanish, a $\U\times\U$-invariant spacetime is called Gowdy spacetime
\cite{Gowdy73}. Thus, all $\U\times\U$-invariant solutions of
\Eqref{eq:EFE} with spatial topology \SoXSt (or \Sth) are Gowdy
solutions.

It is clear that that there exist coordinate gauges which are
inconsistent with the assumption, that for all
$t=const$-hypersurfaces, the first and second fundamental form are
$\U\times\U$-invariant and that the Killing vector fields can be
identified with the coordinate vector fields $\partial_\rho$ and
$\partial_\phi$ for all values of the time coordinate $t$.  Most
prominent examples of -- in this sense -- consistent gauges are the
so-called areal gauge where one sets $\tilde R=t$, and the ``conformal
time gauge''\footnote{The name ``conformal time gauge'' must not be
  confused with the conformal approach described
  in \Sectionref{sec:formulation}. This name was chosen because the
  $2$-surfaces orthogonal to the group orbits are explicitly
  conformally flat.} where the time coordinate
$t$ is chosen so that
\begin{equation}
  \label{eq:ConfGaugemetrik}
  g=-e^{2\lambda}dt^2+h
\end{equation}
with $h$ given by \Eqref{eq:Gowdyh} together with
\Eqsref{eq:GowdyS1xS2behavior}; for more information see for instance
\cite{Clausen07}.  A further ``consistent'' choice of gauge is the
Gauss gauge. Here the time coordinate $t$ is chosen so that metric $g$
takes the form
\begin{equation}
  \label{eq:Gaussmetrik}
  g=-dt^2+h
\end{equation}
with the same $h$ as before.

\paragraph{Spatial homogeneity and the Nariai case}
If $(M,g)$ is spatially homogeneous, it is in particular Gowdy
symmetric, and hence the metric takes the following form in the
conformal time gauge
\begin{equation}
  \label{eq:SpatHom}
  g=\tilde Re^{\tilde P}(-dt^2+d\theta^2
  +\sin^2\theta d\phi^2) +\tilde Re^{-\tilde P}d\rho^2.
\end{equation}
All functions in this metric only depend on $t$. In this gauge, the
generalized Nariai metrics are determined by
\begin{equation}
  \label{eq:NariaiGowd}
  \tilde R(t)=\Phi(t)/\Lambda,\quad \tilde P(t)=-\ln\Phi(t).
\end{equation}


%% file: initial_data.tex
\subsection{A family of Gowdy symmetric initial data close to Nariai data}
\label{sec:initialdata}


Our aim is now to find Gowdy invariant initial data close to the
Nariai solution. These will be interpreted as perturbed Nariai data,
and the corresponding solutions of the field equations as perturbed
Nariai solutions. Our data sets must be solutions of the constraint
equations implied by the vacuum Einstein's field equations with
$\Lambda>0$ on a Cauchy surface of \SoXSt-topology.

Since Gowdy symmetry is generated by the coordinate vector fields
$\partial_\rho$ and $\partial_\phi$, our initial value problem reduces
to a problem on the domain of the coordinates $(t,\theta)$ in
principle.  Let $t$ be the time coordinate of the ``conformal time
gauge'' defined in \Eqref{eq:ConfGaugemetrik}.  The constraints are
\begin{itemize}
\item Hamiltonian constraint: 
  \[0=\frac{1}{4} {P^\prime}^2+\frac{1}{4} e^{2 P}
  {Q^{\prime}}^2-\frac{{R^{\prime}}^2}{4 R^2}+\frac{1}{4} \dot
  P^2+\frac{1}{4} e^{2 P} \dot Q^2-\frac{\dot R^2}{4 R^2}+e^{2
    \lambda } \Lambda -\frac{R^{\prime} \lambda
    ^{\prime}}{R}+\frac{R^{\prime\prime}}{R}-\frac{\dot R
    \dot\lambda}{R},\]
\item Momentum constraint
  \[0=\frac{1}{2} P^{\prime} \dot P+\frac{1}{2} e^{2 P} Q^{\prime}
  \dot Q-\frac{R^{\prime} \dot R}{2 R^2}-\frac{\lambda ^{\prime}
    \dot R}{R}-\frac{R^{\prime} \dot\lambda}{R}+\frac{\dot
    R^\prime}{R}.\]
\end{itemize}
A dot represents a $t$-derivative and a prime a $\theta$-derivative.
We assume that on the initial hypersurface, the action of the Gowdy
group is of the standard form, which implies that the quantities
$R,P,Q,\lambda$ must be expressible via
\Eqsref{eq:GowdyS1xS2behavior}. When the new quantities $\tilde R$,
$\tilde P$ and $\tilde\lambda$ are substituted into the constraint
equations, formally singular terms arise at the coordinate
singularities $\theta=0,\pi$. In this paper, we do not address the
problem of these terms; another future publication will be devoted to
such and related issues. In the case $\Lambda=0$, the corresponding
problem arises \cite{garfinkle1999}. For $\Lambda>0$, however,
Eq.~(14) in \cite{garfinkle1999} must be substituted by a more
complicated condition due to the additional term
$e^{2\tilde\lambda}\Lambda$. In order to circumvent these problems for
the time being as in \cite{garfinkle1999}, we are satisfied with a
particular family of explicit solutions of the constraints with the
properties above for this paper. However, it is clear that such a
family of initial data cannot be considered ``generic'' and hence no
strict results about the cosmic no-hair conjecture can be
expected. Nevertheless, we see the investigations in this paper as a
promising first step.

In all of what follows, we assume $\Lambda=3$ with loss of generality,
since it yields the simplest expressions and makes all quantities
dimensionless.  We have mentioned before that the functions $\tilde
R$, $\tilde P$, $Q$ and $\tilde\lambda$ must be smooth functions of
$\cos\theta$, and $\tilde\lambda$ has to become zero at
$\theta=0,\pi$. Now, we make a polynomial ansatz for these functions
in $z=\cos\theta$, and solve the constraints for the polynomial
coefficients matching these conditions. In this way, which requires
cumbersome algebra done with Mathematica, we derive the following
almost explicit family of Gowdy symmetric solutions of the constraints
\begin{subequations}
  \label{eq:familyID}
  \begin{align}
    \tilde R&=\tilde R_*,& \tilde R^\prime&=\frac{\tilde R_*}{\kappa},\\
    \tilde P&=P_*-\frac{\sqrt 3}{2\kappa}\tilde N_\times^{(1)}\sin^2\theta,&    
    \tilde P^\prime&=\frac{\sqrt{3}}\kappa(
    \Sigma_-^{(0)}-\Sigma_\times^{(1)}\cos^2\theta),\\
    Q&=-\frac{\sqrt{3}}\kappa\tilde N_\times^{(1)}
    \int_{-1}^{\cos\theta} e^{-\tilde P(z)}dz,&
    Q^\prime&=\frac{\sqrt{3}}\kappa \Sigma_\times^{(1)} e^{-\tilde P},\\
    \tilde\lambda&=\frac{\sqrt{3}}{4\kappa}
    \tilde N_\times^{(1)}\sin^2\theta,&
    \tilde\lambda^\prime&=\frac{\sqrt{3}}{2\kappa}(
    \sqrt3\Sigma_+^{(2)}-\Sigma_\times^{(1)})\sin^2\theta.
  \end{align}
\end{subequations}
The constants $P_*$, $\tilde N_\times^{(1)}$, $\Sigma_+^{(2)}$ are
determined transcendentally by the initial data parameters $(\tilde
R_*,\kappa, \Sigma_-^{(0)}, \Sigma_\times^{(1)})$ in the following
manner. Let us make the abbreviation
\begin{equation*}
  C:=\frac{1}{\sqrt{3}\,\kappa}
  \sqrt{\kappa ^4
    +\left(3 (\Sigma_\times^{(1)})^2-6 \Sigma_-^{(0)} \Sigma_\times^{(1)}
      +2 \sqrt{3}\Sigma_-^{(0)}\right) \kappa ^2
    +3 (\Sigma_\times^{(1)}-\Sigma_-^{(0)})^2}.
\end{equation*}
Then 
\begin{align*}
  P_*&=\ln\left[\frac{1}{4\tilde R_*\kappa^2}
    \left(
      \frac{2 \kappa ^2}{3}-\frac{2 C \kappa }{\sqrt{3}}
      -(\Sigma_\times^{(1)})^2-(\Sigma_-^{(0)})^2
      +2 \Sigma_\times^{(1)} \Sigma_-^{(0)}+1
    \right)\right],\\
  \tilde N_\times^{(1)}&=\frac{\Sigma_\times^{(1)}-\Sigma_-^{(0)}}{\kappa }
  -C-\frac{\kappa}{\sqrt{3}},\\
  \Sigma_+^{(2)}&=\frac{3 (\Sigma_\times^{(1)})^2
    -6 \Sigma_-^{(0)}\Sigma_\times^{(1)}
    +3 C \kappa(\Sigma_-^{(0)}-\Sigma_\times^{(1)})
    +\Sigma_-^{(0)} \left(\sqrt{3} \kappa^2+3 \Sigma_-^{(0)}\right)}
  {3 \kappa ^2}.
\end{align*}
The reality conditions on both the root in the definition of $C$ and
the logarithm in the definition of $P_*$ imply restrictions for the
choice of the (otherwise free) parameters $(\tilde R_*,\kappa,
\Sigma_-^{(0)}, \Sigma_\times^{(1)})$ which, however, we do not make
explicit now.  For the applications later, we always check that these
are satisfied without further notice.  Let us suppose that all
quantities in \Eqsref{eq:familyID} are well defined. The only
non-explicit expression is the integral for $Q$ when
$\Sigma_\times^{(1)}\not=0$. We compute this integral numerically by
approximating the exponential by its truncated Taylor series. This
series converges very quickly and in practice, the series can be
truncated after a few terms.

Let us also remark that these data are
not polarized in general, i.e.\ the Killing fields cannot be chosen
globally orthogonal.

Now, let us identify spatially homogeneous, and in particular Nariai
data in our family \Eqref{eq:familyID}.  Writing \Eqref{eq:SpatHom}
for the conformal time gauge according to \Eqref{eq:ConfGaugemetrik},
we see that spatial homogeneity implies $\Sigma_\times^{(1)}=\tilde
N_\times^{(1)}=\Sigma_+^{(2)}=0$.  Using the expressions above, the
data are hence spatially homogeneous if and only if
\begin{equation}
  \label{eq:SpatHomCond}
  \Sigma_-^{(0)}\le-\frac{\kappa^2}{\sqrt 3}, \quad \Sigma_\times^{(1)}=0.
\end{equation}
In particular, $\Sigma_\times^{(1)}$ plays the role of an
``inhomogeneity parameter''.  However, our family of data does not
comprise all spatially homogeneous data, in particular not all Nariai
data. For any generalized Nariai data, \Eqref{eq:NariaiGowd} implies
\begin{equation}
  \label{eq:relationNariai}
  \tilde R_*=\Phi_*/\Lambda,\quad 
  \kappa=\Phi_*/\Phi_*^\prime,\quad \Sigma_-^{(0)}=-1/\sqrt{3},\quad
  \Sigma_\times^{(1)}=0.
\end{equation}
In this case, \Eqref{eq:SpatHomCond} yields $\kappa^2\le 1$, and thus
only generalized Nariai solutions with $\sigma_0\le 0$ are present in
our data.


%% file: expansions.tex
\subsection{Mean curvatures and the expected instability in the Gowdy class}
\label{sec:expansions}
In principle, there is no ``canonical'' quantity which could play the
same role for the instability of the Nariai solutions in the spatially
inhomogeneous case as the quantity $H^{(0)}_*$ defined
in \Sectionref{P1-sec:ansatzKS} in the first paper
\cite{beyer09:Nariai1} in the spatially homogeneous case.  First,
there is no ``canonical'' foliation of spacetime, and, second, no
``geometrically preferred spatial \St-factor''.  Here, we choose a
Gaussian foliation with time coordinate $t$; cf.\
\Eqref{eq:Gaussmetrik}. At this stage, we can only hope that at least
for small perturbations of the Nariai solutions, the ``expansion of
the coordinate \St-factor'', which we define now, plays a similar role
here as $H^{(0)}_*$; at least these quantities agree for spatially
homogeneous perturbations.

On the initial hypersurface, we choose coordinates
$(\rho,\theta,\phi)$ as before. By means of the Gauss gauge condition,
these spatial coordinates are transported to all
$t=const$-hypersurfaces $\Sigma_t$. On any $\Sigma_t$, a ``coordinate
\St-factor'' is then a $2$-surface diffeomorphic to \St determined by
$\rho=const$. Since the metric is invariant under translation along
$\rho$, all such $2$-surfaces are isometric at a given $t$. Similarly,
we define ``coordinate \So-factors''; note that these $1$-surfaces are
not isometric on a given $\Sigma_t$.  The expansions (mean curvatures)
associated with these surfaces are defined as follows for a Gauss
gauge. Let $H$ be the mean curvature of any
$t=const$-hypersurface. Let $H_2$ be the projection of the mean
curvature vector of a coordinate \St-factor to $\partial_t$; this is
the quantity we refer to as the ``expansion of the coordinate
\St-factor''. Similarly, we define $H_1$ as the ``expansion of the
coordinate $\So$-factor''. With the expression \Eqref{eq:Gowdyh} for
the spatial metric, the following formulas hold
\begin{subequations}
  \label{eq:GowdyS1xS2meancurvatures}
  \begin{gather}
    H=
    \frac{3{\tilde R}^{\prime}
      +{\tilde R}\left({\tilde P}^{\prime}+2 {\tilde\lambda}^{\prime}\right)}
    {6 {\tilde R}},\quad
    H_2=
    \frac{{\tilde R}^{\prime}
      +{\tilde R}\left({\tilde P}^{\prime}+{\tilde\lambda}^{\prime}\right)}
    {2 {\tilde R}},\\
    3 H-2 H_2-H_1=-\frac{e^{2 {\tilde P}} Q \sin^2\theta \left(Q
        {\tilde P}^{\prime}+Q^{\prime}\right)}{1+e^{2 {\tilde P}}Q^2
      \sin^2\theta},
  \end{gather}
\end{subequations}
where a prime denotes a derivative with respect to Gaussian
time\footnote{Recall that, by contrast, a prime denotes a derivative
  with respect to the time coordinate in conformal time gauge in
  \Eqsref{eq:familyID}.}.

We expect that, as for the quantity $H^{(0)}_*$ in the spatial
homogeneous case, the sign of the initial value of $H_2$ controls
whether the solution collapses or expands locally in space and hence
plays a particularly important role for the description of the
expected instability of the Nariai solutions. Thus we write down
the expression for the family of initial data in \Eqsref{eq:familyID}
\begin{equation}
  \label{eq:IDH2}
  \begin{split}
    \left.H_2\right|_{\text{initial}}
    =&\frac1{4\kappa^3}\Bigl[3 (\Sigma_\times^{(1)}-\Sigma_-^{(0)})^2
    -3\kappa C (\Sigma_\times^{(1)}-\Sigma_-^{(0)})
    +\kappa^2(2-\sqrt3\Sigma_\times^{(1)}+3\sqrt 3\Sigma_-^{(0)})\\
    &-\Bigl(3 (\Sigma_\times^{(1)}-\Sigma_-^{(0)})^2
    -3\kappa C (\Sigma_\times^{(1)}-\Sigma_-^{(0)})
    +\sqrt3\kappa^2(\Sigma_\times^{(1)}+\Sigma_-^{(0)}\Bigr)\cos^2\theta\Bigl].
  \end{split}
\end{equation}


%% file: formulation_numerics.tex
\subsection{Formulation of Einstein's field equations}
\label{sec:formulation}

Our numerical approach, based on orthonormal frames for the conformal
field equations is discussed in \cite{beyer08:code,beyer08:TaubNUT}
and is briefly summarized in the following paragraphs.  The main
motivation for using this approach is that the conformal techniques
allow us, in principle, to compute the conformally extended solutions
including conformal boundaries. Recall that in our setting, smooth
conformal boundaries represent the infinite timelike future or past,
and hence play a particular role for the cosmic no-hair picture;
cf.\ \Sectionref{P1-sec:cosmicnohair} in the first paper
\cite{beyer09:Nariai1}.

The ``physical metric'', by which we mean a solution of Einstein's
field equations, is denoted by $\tilde g$, and all
corresponding quantities (connection coefficients, curvature tensor
components etc.) are marked with a tilde%
\opt{longversion}{%
  \footnote{Be aware that this is the opposite
    notation than in \cite{beyer09:Nariai1}; there, all quantities
    defined with respect to the conformal metric are marked with a
    tilde.}%
}%
.  The so-called ``conformal metric'' on the conformal
compactification is denoted by $g$; all corresponding quantities are
written without a tilde. Both metrics are related by the expression
$g=\Omega^2\tilde g$, where $\Omega>0$ is a conformal factor. We
cannot give further explanations here; some more details are listed in
the first paper \cite{beyer09:Nariai1}, and a comprehensive review is
in \cite{Friedrich2002}.

We will use Friedrich's general conformal field equations in a special
conformal Gauss gauge
\cite{AntiDeSitter,Friedrich2002,Lubbe09,beyer:PhD,beyer08:TaubNUT}. In
turns out that up to a rescaling of the time coordinate $t$ of this
gauge, it is equivalent to a physical Gauss gauge (defined with
respect to $\tilde g$) with time coordinate $\tilde t$. For
$\Lambda=3$, the rescaling has the form
\begin{equation*}
  \tilde t=\ln\frac t{2-t}.
\end{equation*}
If a smooth compact past conformal boundary \scrim exists, and if the
solution extends to the conformal boundary in this gauge, then \scrim
equals the $t=0$-hypersurface where $\tilde t\rightarrow
-\infty$. Under analogous conditions, \scrip is represented by the
$t=2$-hypersurface where $\tilde t\rightarrow\infty$.

\begin{subequations}
  \label{eq:orth_frame_Ya}
  Now we write the evolution equations and list the unknowns. We always
  assume $\Lambda=3$ in order to obtain the simplest expressions as
  possible. Among the unknown fields is a smooth frame $\{e_i\}$, which
  is orthonormal with respect to $g$, and which we represent as
  follows. Due to our gauge choice, we can fix
  \begin{equation}
    e_0=\partial_t 
  \end{equation}
  which is henceforth the future directed unit
  normal, with respect to $g$, of the $t=const$-hypersurfaces.
  Furthermore, we write
  \begin{equation}    
    e_a=e\indices{_a^b}V_b,
  \end{equation}
\end{subequations}
where $(e\indices{_a^b})$ is a smooth $3\times 3$-matrix valued
function with non-vanishing determinant on \SoXSt. Let us define
\begin{equation}
  \label{eq:KS_KillingBasis}
  W_1=\sin\phi\partial_\theta+\cos\phi\cot\theta\partial_\phi,\quad
  W_2=\cos\phi\partial_\theta-\sin\phi\cot\theta\partial_\phi,\quad
  W_3=\partial_\phi,      
\end{equation}
and from those the vector fields
\begin{equation}
  \label{eq:frameS1xS2}
  \begin{split}
    V_1&=2(-\sin\theta\cos\phi\, \partial_\rho+W_1),\quad
    V_2=2(\sin\theta\sin\phi\, \partial_\rho+W_2),\\
    V_3&=2(\cos\theta\, \partial_\rho+W_3).
  \end{split}
\end{equation}
The factors $2$ are chosen for later convenience.  It turns out that
$\{V_a\}$ forms a smooth global frame on \SoXSt.

Having fixed the residual gauge initial data, as described in
\cite{beyer:PhD}, a hyperbolic reduction of the general conformal
field equations is given by
\begin{subequations}%
  \label{eq:gcfe_levi_cevita_evolution}%
  \begin{align}
    \label{eq:evolution_frame}
    \partial_t e\indices{_a^c}&=-\chi\indices{_a^b}e\indices{_b^c},\\
    \partial_t\chi_{ab}
    &=-\chi\indices{_{a}^c}\chi_{cb}
    -\Omega E_{ab}
    +L\indices{_{{a}}_{b}},\\
    \partial_t\Connection abc
    &=-\chi\indices{_a^d}\Connection dbc
    +\Omega B_{ad}\epsilon\indices{^b_c^d},\\
    \partial_t L_{ab}
    &=-\partial_t\Omega\, E_{ab}-\chi\indices{_a^c}L_{cb},\\
    \label{eq:Bianchi1}
    \partial_t E_{fe}-D_{e_c}B_{a(f}\epsilon\indices{^a^c_{e)}}
    &=-2\chi\indices{_c^c}E_{fe}
    +3\chi\indices{_{(e}^c}E_{f)c}
    -\chi\indices{_c^b}E\indices{_b^c}g_{ef},\\
    \label{eq:Bianchi2}
    \partial_t B_{fe}+D_{e_c}E_{a(f}\epsilon\indices{^a^c_{e)}}
    &=-2\chi\indices{_c^c}B_{fe}
    +3\chi\indices{_{(e}^c}B_{f)c}
    -\chi\indices{_c^b}B\indices{_b^c}g_{ef},\\
    \label{eq:conffactor}
    \Omega(t)&=\frac 12\, t\, (2-t),
  \end{align}
  for the unknowns 
  \begin{equation}
    \label{eq:unknowns}
    u=\left(e\indices{_a^b}, \chi_{ab}, \Connection abc, L_{ab}, E_{fe},
      B_{fe}\right).
  \end{equation}
\end{subequations}
The unknowns $u$ are the spatial components $e\indices{_a^b}$ of a
smooth frame field $\{e_i\}$ as in \Eqref{eq:orth_frame_Ya}, the
spatial frame components of the second fundamental form $\chi_{ab}$
defined with respect to $e_0$, the spatial connection coefficients
$\Connection abc$, given by $\Connection abc
e_b=\nabla_{e_a}e_c-\chi_{ac}e_0$ where $\nabla$ is the Levi-Civita
covariant derivative operator of the conformal metric $g$, the spatial
frame components of the Schouten tensor $L_{ab}$, which is related to
the Ricci tensor of the conformal metric by
\begin{equation*}
L_{\mu\nu}=R_{\mu\nu}/2
  -g_{\mu\nu}g^{\rho\sigma}R_{\rho\sigma}/12,
\end{equation*}
and the spatial frame components of the electric and magnetic parts of
the rescaled conformal Weyl tensor $E_{ab}$ and $B_{ab}$
\cite{Friedrich2002,FriedrichNagy}, defined with respect to $e_0$.
Because the timelike frame field $e_0$ is hypersurface orthogonal,
$\chi_{ab}$ is a symmetric tensor field. In order to avoid confusion,
we point out that, in general, the conformal factor $\Omega$ is part
of the unknowns in Friedrich's formulation of the CFE. However, for
vacuum with arbitrary $\Lambda$ and for arbitrary conformal Gauss
gauges, it is possible to integrate its evolution equation explicitly
\cite{AntiDeSitter}, so that $\Omega$ takes the explicit form
\Eqref{eq:conffactor} for our choice of gauge.  We note that, $E_{ab}$
and $B_{ab}$ are tracefree by definition. Hence we can get rid of one
of the components of each tensor, for instance by substituting
$E_{33}=-E_{11}-E_{22}$; we do the same for the magnetic part. The
evolution equations \Eqsref{eq:Bianchi1} and \eqref{eq:Bianchi2} of
$E_{ab}$ and $B_{ab}$ are derived from the Bianchi system
\cite{Friedrich2002}. In our gauge, the constraint equations implied
by the Bianchi system take the form
\begin{equation}
  \label{eq:bianchi_constraints}
  D_{e_c} E\indices{^c_e}
  -\epsilon\indices{^a^b_e}B_{da}\chi\indices{_b^d}=0,\quad
  D_{e_c} B\indices{^c_e}
  +\epsilon\indices{^a^b_e}E_{da}\chi\indices{_b^d}=0.
\end{equation}
Here, $\epsilon\indices{_a_b_c}$ is the totally antisymmetric symbol
with $\epsilon\indices{_1_2_3}=1$, and indices are shifted by means of
the conformal metric.  The other constraints of the full system above
are equally important, but are ignored for the presentation here.
Note that in \Eqsref{eq:Bianchi1}, \eqref{eq:Bianchi2} and
\eqref{eq:bianchi_constraints}, the fields $\{e_a\}$ are henceforth
considered as spatial differential operators, using
\Eqref{eq:orth_frame_Ya} and writing the fields $\{V_a\}$ as
differential operators in terms of coordinates according to
\Eqref{eq:frameS1xS2} and \eqref{eq:KS_KillingBasis}. Interpreted as
partial differential equations, these evolution equations are
symmetric hyperbolic and the initial value problem is
well-posed. Further discussions of the above evolution system and the
quantities involved can be found in the references above.

Friedrich's CFE allow us to use \scrip, i.e.\ the $t=2$-surface (or in
the same way \scrim), as the initial hypersurface. This particular
initial value problem was considered in
\cite{beyer08:TaubNUT}. However, in our present application, not all
solutions of interest have smooth conformal boundaries. In order not
to exclude those solutions, we choose the $t=1$-hypersurface as the
initial hypersurface, which is a standard Cauchy surface. The hope is
that this setup allows us to compute the complete solution including
the conformal boundary if it exists.

It is a standard result that there exist no globally smooth frames on
\SoXSt with the property that each frame vector field has vanishing
Lie brackets with both Gowdy Killing vector fields $\partial_\rho$ and
$\partial_\phi$. The reason is given for instance in
\cite{beyer08:code}. It is only possible to find a frame whose Lie
brackets vanish for one of the two Killing vector fields, say,
$\partial_\rho$. This, however, has the consequence that the frame
components of all tensor fields derived from a Gowdy invariant metric
$g$ depend on the coordinate $\phi$ in a non-trivial manner. In order
to reduce the evolution equations based on such a frame to $1+1$
dimensions nevertheless, we can do the following \cite{beyer08:code}.
It turns out to be possible to evaluate the $\phi$-derivative of every
relevant unknown at, say, $\phi=0$ algebraically in terms of the
unknowns. Then one can write an evolution system which only involves
$\theta$-derivatives in space by substituting all $\phi$-derivatives
with these algebraic expressions. The resulting system, which we
called $1+1$-system in \cite{beyer08:code} in the case of spatial
\Sth-topology, is symmetric hyperbolic for the conformal field
equations in our gauge.
\opt{longversion}{%
  It follows from the discussion in \Sectionref{sec:actionU1xU1S1xS2S3},
  that a similar reduction to $1+1$ is possible for spatial
  \SoXSt-topology. The resulting system of equations is used exclusively
  in all of what follows in this paper.
}%
%
%
\opt{shortversion}{%
  It turns out that a similar reduction to $1+1$ is possible for spatial
  \SoXSt-topology. The resulting system of equations is used exclusively
  in all of what follows in this paper.
}%

\subsection{Numerical infrastructure}
\label{sec:numerics}
\opt{longversion}{%
  It follows from \Sectionref{sec:actionU1xU1S1xS2S3} that for Gowdy
  symmetry, spatial \SoXSt- and \Sth-topology have the same
  representation. 
  Basically, we only need to substitute the Euler coordinates
  $(\chi,\rho_1,\rho_2)$ in the \Sth-case defined in
  \Eqsref{eq:eulerangleparm} and \eqref{eq:defrho} by the coordinates
  $(\theta,\phi,\rho)$ in the \SoXSt-case, and the reference frame
  $\{Y_a\}$ defined in \Eqsref{eq:S3referencefields} by the $\{V_a\}$
  mentioned before.  Then, it is possible to use the same numerical
  technique as that which was worked out originally for spatial
  \Sth-topology in \cite{beyer08:code}.

}%
%
%
\opt{shortversion}{%
  Thanks to the similarity of the manifolds \SoXSt- and \Sth from the
  point of view of Gowdy symmetry, only minimal changes to the
  \Sth-code presented in \cite{beyer08:code} are necessary. We will
  not go into these details.  
}%
Let us repeat quickly the main ingredients of this code.  By means of
the Euler coordinates of \Sth, it is possible transport all geometric
quantities and hence Einstein's field equations themselves from \Sth
to \T; loosely speaking, we make ``all spatial directions
periodic''. It is clear that such a map must be singular at some
places. However, it is possible to analyze the behavior of Fourier
series at the singular places and to compute the formally singular
terms in the equations explicitly. Hence, it is not only natural to
use Fourier based pseudospectral spatial discretization due to the
periodicity in each spatial direction on \T here, but it also allows
to regularize the formally singular terms in spectral space. This is
the motivation for choosing spectral discretization in
space. Nevertheless, a scheme to enforce ``boundary conditions''
\cite{beyer08:code} can be necessary in practice to guarantee the
numerical smoothness and stability. We come back to this when we
present our results in \Sectionref{sec:numresults}.

For the time discretization, we use the method of lines. In this work,
all numerical results were obtained with the adaptive $5$th-order
``embedded'' Runge Kutta scheme from \cite{numericalrecipes} unless
noted otherwise. 

\subsection{Numerical computation of the initial data for the CFE}
\label{sec:numcompID}
In \Sectionref{sec:initialdata}, we constructed initial data for the
functions $\tilde R$, $\tilde P$, $Q$ and $\tilde\lambda$. Now we fix
the initial value of the frame, i.e.\ the components $e\indices{_a^b}$
in \Eqref{eq:orth_frame_Ya}, and compute the corresponding initial
values of $u$ in \Eqref{eq:unknowns}. 

We choose the initial value of $e\indices{_a^b}$ as follows. For the
family of initial data constructed in \Sectionref{sec:initialdata},
consider the frame $\{V_a\}$ in \Eqref{eq:frameS1xS2} and perform a
Gram-Schmidt orthonormalization with respect to the initial conformal
$3$-metric. More precisely, we construct the matrix
$(e\indices{_a^b})$ from \Eqref{eq:orth_frame_Ya} as an upper
triangular matrix. For instance, this means that $e_3$ and $V_3$ are
collinear initially. Of course there is a great freedom of choosing
frames, and this choice is just one possibility. Note that for the
following, no time derivative of $e\indices{_a^b}$ at the initial time
$t=1$ needs to be prescribed.

Now we comment on the computation of $u$ \Eqref{eq:unknowns} from
these data. The data in \Sectionref{sec:initialdata} yield all spatial
derivatives of the initial metric components and the first time
derivatives.  However, in order to compute $u$ at the initial time
$t=1$ from these data, we also need second time derivatives of the
data.  We calculate these by imposing the evolution equations of
Einstein's field equations at $t=1$. We decided to perform all the
computations numerically.  Note that for this, the metric functions of
the initial data in \Sectionref{sec:initialdata} yield formally
singular terms at $\theta=0,\pi$. However, we are able to compute
these formally singular terms numerically by applying the spectral
approach which was described above in the context of the evolution
equations. In practice, we find that this allows us to resolve the
data $u$ with high accuracy. It turns out that machine round-off
errors, i.e.\ errors introduced by the finite number representation in
the computer, often yield the largest error contributions here.  This
is true in particular when the standard ``double precision'' with
round-off errors of order $10^{-16}$ on Intel processors is used.
Hence, we decided to compute the initial data with ``quad precision''
of the Intel Fortran compiler \cite{Intel}, where numbers are
represented with roughly $32$ digits, but which is software emulated
and hence relatively slow. For the evolution, we switch back to double
precision. All our numerical computations presented here have been
obtained in this way.


%% file: numresults_smallamplitudes.tex
\subsection{Choice of perturbed data}
\label{sec:choicedata}
Let us proceed by explaining our particular choices of initial data
for the following numerical results. We present only perturbations of
one generalized Nariai solution given by $\tilde R_*=1.0$,
$\kappa=0.5$ here; cf.\ \Eqref{eq:relationNariai}.  In order to
perturb these Nariai data in an interesting manner, we choose the
inhomogeneity parameter $\Sigma_\times^{(1)}$ in \Eqref{eq:familyID}
as non-zero, but small, and furthermore introduce a small non-zero
parameter $\mu$ by
\[\Sigma_-^{(0)}=-1/\sqrt 3+\mu.\]
For the value $\mu=\Sigma_\times^{(1)}=0$, the data reduce to Nariai
data according to \Eqref{eq:relationNariai}. In all of what follows,
we show the numerical results for three initial data sets given by
\[\mu=0.0004667, \quad 0.0004800, \quad 0.0005000,\]
and $\Sigma_\times^{(1)}=4\cdot 10^{-4}$ in all these three cases.
These choices are motivated as follows.  On the one hand, we want to
focus on ``small perturbations'' as a first step in order to study the
instability of the Nariai solution carefully. We believe that the
choice of the parameters above is consistent with this. Indeed, we
have experimented with other values of these parameters. In particular
in a large range of values for $\Sigma_\times^{(1)}$, there is no
qualitative change in the results that follow. When we go to ``very
large'' values $\Sigma_\times^{(1)}\sim 10^{-1}$, then a different
phenomenology occurs; this interesting aspect is currently under
investigation and will not be presented in this paper.  When we,
however, go to even smaller values of $\Sigma_\times^{(1)}$, we get
problems with numerical accuracy. Numerical errors in our runs are
discussed in
\Sectionref{sec:DetailsErrors}. We remark that in this case of small
perturbations, a linearization of the field equations around the
unperturbed Nariai solution can be a reasonable approximation. This is
currently work in progress, but will not be presented here; indeed,
all numerical results that follow are based on the full non-linear
field equations.

In any case, smallness of the parameters is not the only motivation
for our particular choices of data above. Recall that we expect that
the instability of the Nariai solution can be exploited to construct
cosmological black hole solutions. Our expectation is that this
instability is controlled by the sign of the initial value of $H_2$.
\begin{figure}[t]
  \centering
  \psfrag{theta}[][][0.9]{$\theta$}
  \psfrag{H2}[][][0.9]{$H_2$}
  \includegraphics[width=0.49\textwidth]{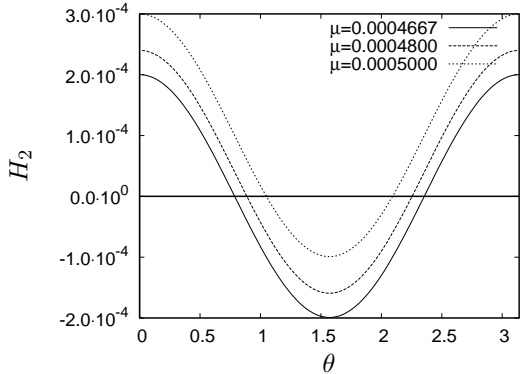}
  \caption{Initial spatial dependence of $H_2$ for the three initial
    data sets considered here.}
  \label{fig:initial_HH2_1D}
\end{figure}
For our choices of parameters, $H_2$ has positive and negative parts
according to \Eqref{eq:IDH2}, see \Figref{fig:initial_HH2_1D}.  Our
expectation for these data sets is hence that in the future, all these
solutions collapse and form the interior of a cosmological black hole
solution at those spatial places where $H_2<0$ initially, i.e.\ close
to the equator of the spatial \St-factor, and expand and form the
cosmological region with a smooth piece of \scrip at those spatial
places where $H_2>0$ initially, i.e.\ at the poles of the spatial
\St-factor. Changing the value of $\mu$ with fixed
$\Sigma_\times^{(1)}$ shifts the initial spatial profiles of $H_2$
``vertically'' in \Figref{fig:initial_HH2_1D} leaving the shape and
the amplitude of the curves approximately invariant. Due to this, we
expect that the larger $\mu$ is, the ``smaller'' should be the black
hole region of the resulting solution.

\subsection{The numerical results}
\label{sec:thenumresults}
Now we present our numerical results based on these data sets;
practical details and a discussion of numerical errors are given
afterwards in \Sectionref{sec:DetailsErrors}.

\paragraph{Future evolution}
\begin{figure}[t]
  \centering
  \psfrag{t}[][][0.9]{$t$}
  \psfrag{Max(H2)}[][][0.9]{$\text{max}(H_2)$}
  \psfrag{Min(H2)}[][][0.9]{$\text{min}(H_2)$}
  \includegraphics[width=0.49\textwidth]{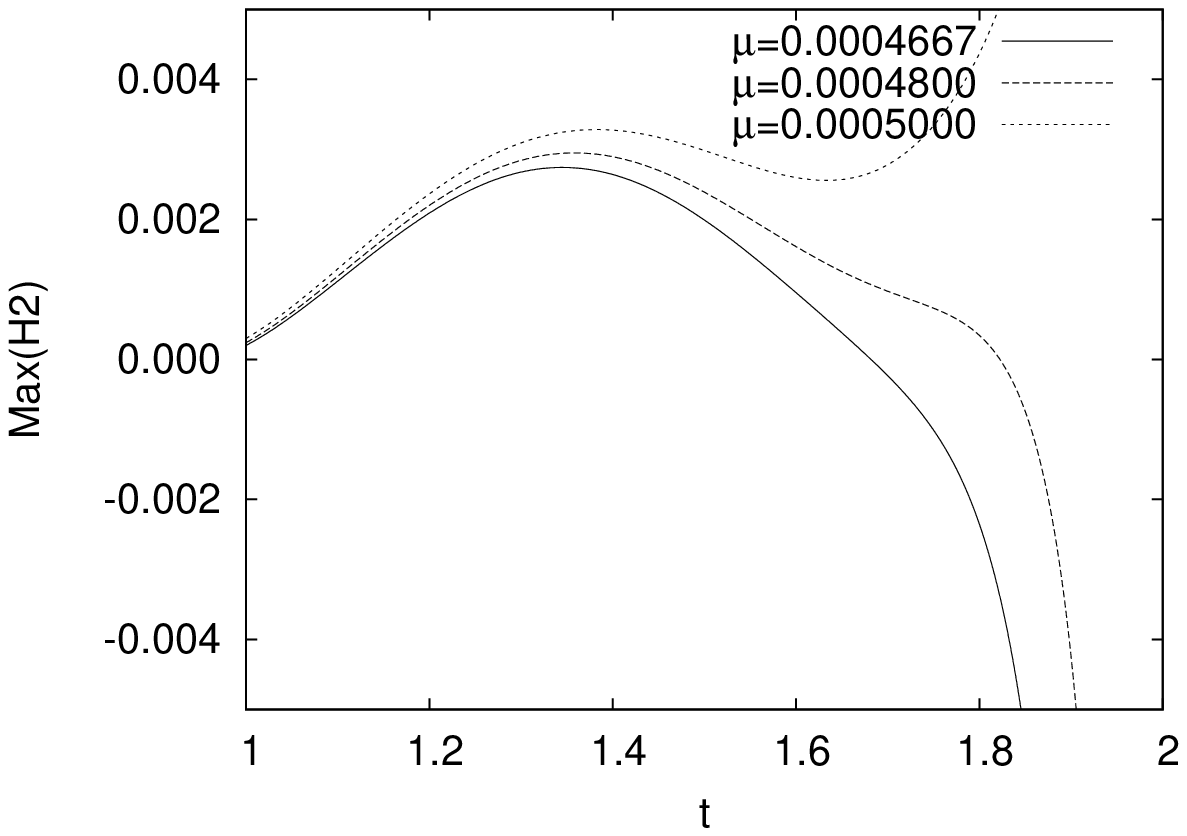}
  \includegraphics[width=0.49\textwidth]{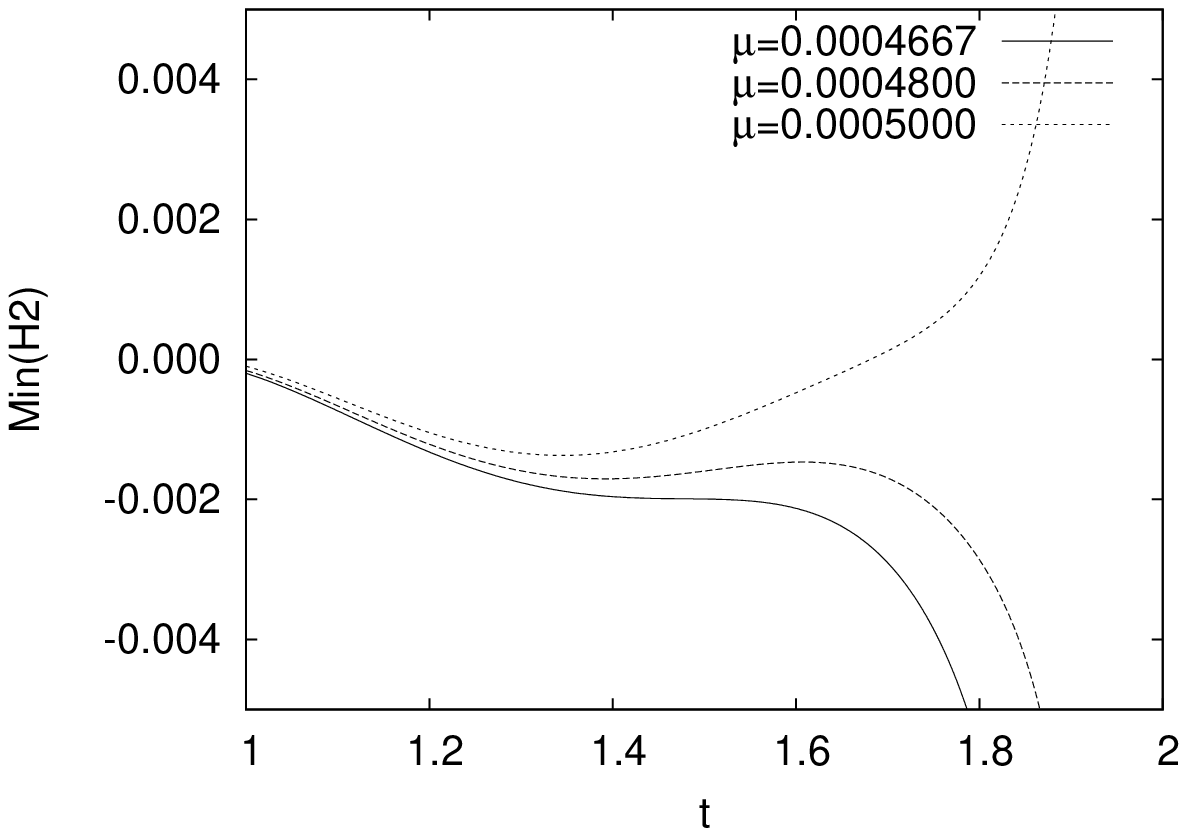}
  \caption{Spatial minimum and maximum of $H_2$ vs.\ time.}
  \label{fig:minmaxH2}
\end{figure}
The horizontal axes in the plots in \Figref{fig:minmaxH2} represent
the time coordinate $t$; recall that the initial hypersurface
corresponds to $t=1$, and $t=2$ would correspond to the infinite
timelike future. Hence, these plots show the future evolution of $H_2$
for our three initial data sets.  On the vertical axis, we show the
maximum and minimum values, respectively, of $H_2$ at any given time
$t$.  In the early phase of the evolution, the solution behaves in
accordance with our expectations.  Basically, the expansion of the
coordinate \St-factor given by $H_2$ becomes more and more positive
where it is positive initially, namely at the maximum at the poles of
the $2$-sphere, see again \Figref{fig:initial_HH2_1D}. Furthermore, it
becomes more and more negative where it is negative initially, namely
at the minimum at the equator of \St. However, at a time $t\approx
1.3$, the behavior changes completely. The spatial profiles of $H_2$
start to become ``flatter'' in the sense that the maximal value of
$H_2$ becomes smaller and the minimal value larger with increasing
$t$, as can be seen from the figure. Eventually the solutions ``make a
decision'' whether the coordinate \St-factor expands or collapses
indefinitely \textit{globally} in space. We give more evidence for
this in a moment.

We do not understand the mechanism underlying this phenomenon yet.  We
hope to be able to to shed further light on this by means of the
linearization of the problem in future work.  In any case, the
numerical results suggest that there is a new instability and a new
critical solution, in addition to the expected instability of the
Nariai solutions. That is, there must be a critical value $\mu_c$ of
$\mu$ in the interval $(0.00048,0.0005)$. For $\mu<\mu_c$, the
solution collapses eventually, and for $\mu>\mu_c$, expands globally
in space. It would be interesting research to identify the
critical solution and to study whether critical phenomena, which play
such an important role \cite{Gundlach2007} for the critical collapse
of black holes, also occur here. In any case, it is an interesting
unexpected result that it does not seem possible to construct
cosmological black hole solutions for small Gowdy symmetric
perturbations of the Nariai solution, in contrast to the claims in
\cite{Bousso03} for the spherically symmetric case.

\begin{figure}[t]
  \centering
  \psfrag{t}[][][0.9]{$t$}
  \psfrag{Max(Kretschmann)}[][][0.9]{$\text{max}(K)$}
  \psfrag{Min(Kretschmann)}[][][0.9]{$\text{min}(K)$}
  \includegraphics[width=0.49\textwidth]{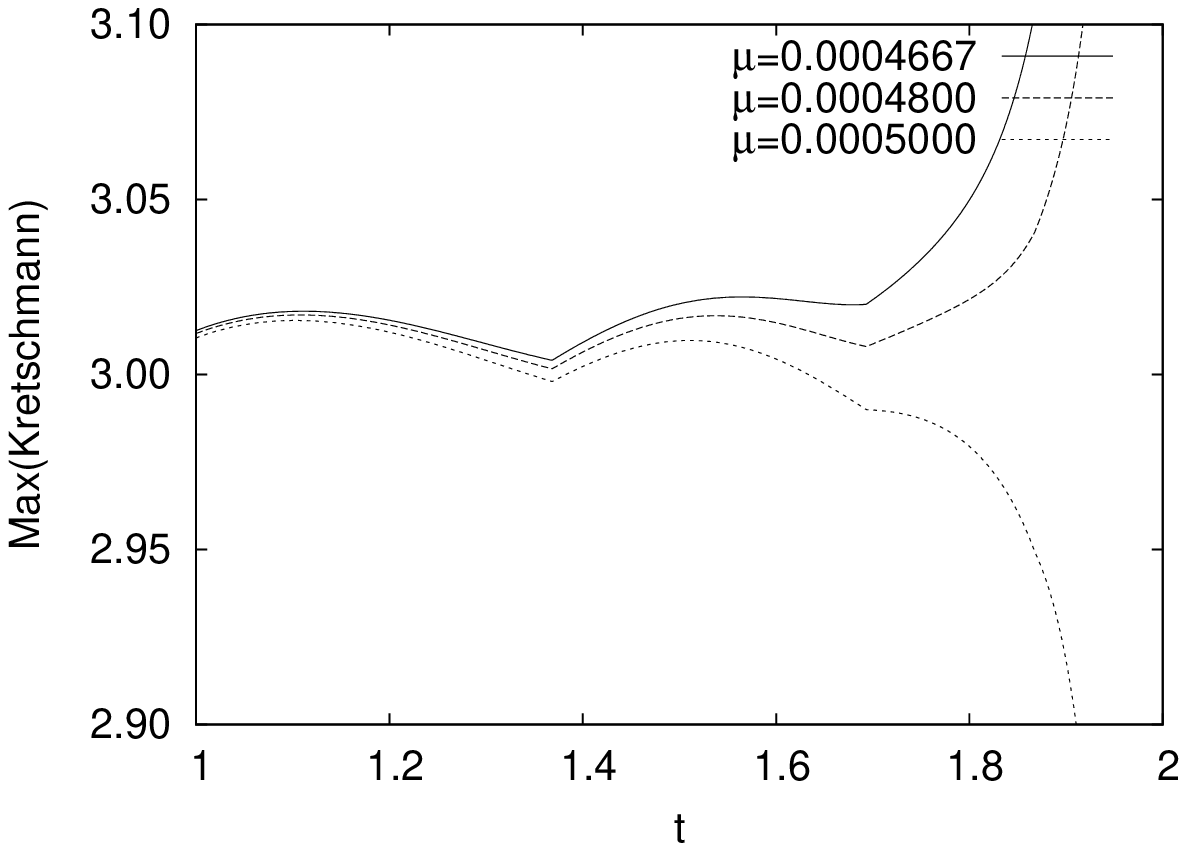}
  \includegraphics[width=0.49\textwidth]{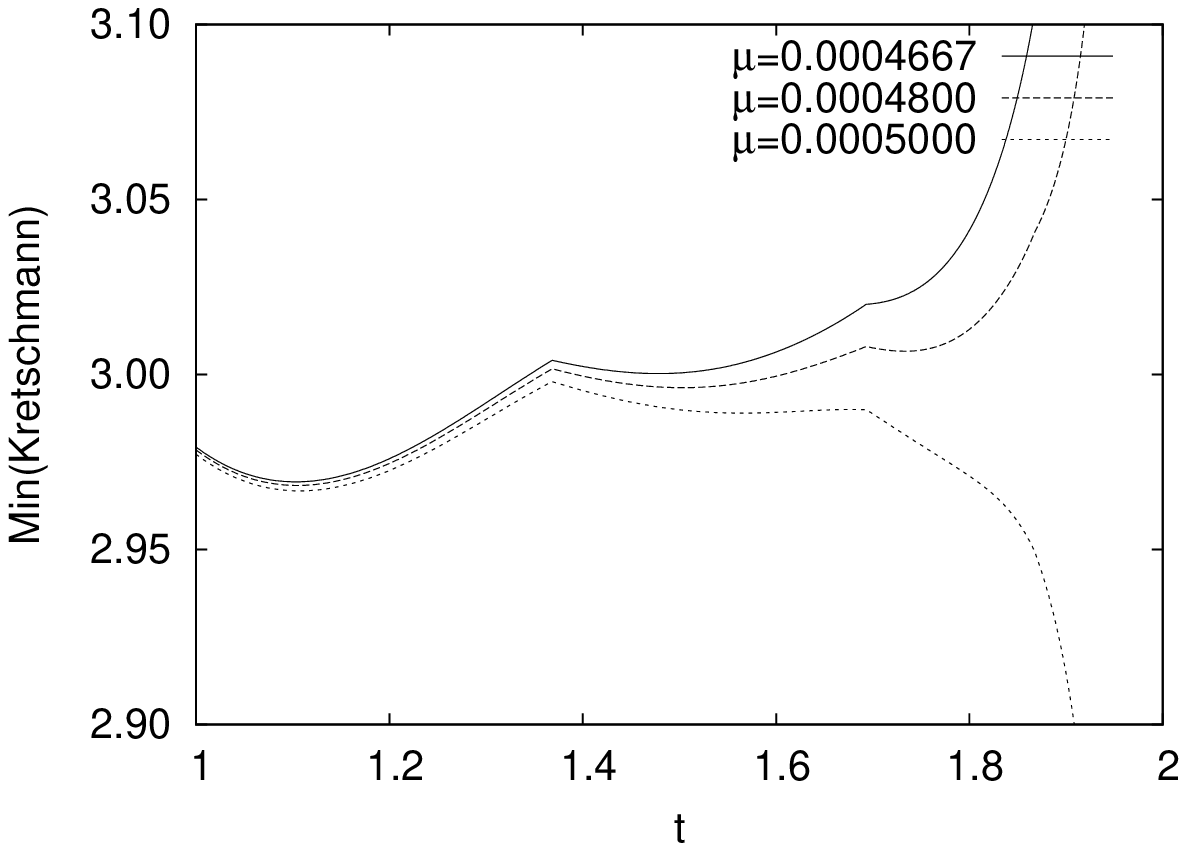}
  \caption{Spatial minimum and maximum the Kretschmann scalar $K$ vs.\
    time.}
  \label{fig:minmaxKretschm}
\end{figure}
We present further evidence for our interpretation of the numerical
results now. For this, consider the plots in
\Figref{fig:minmaxKretschm} for the Kretschmann scalar
\[K:=\tilde R_{\mu\nu\rho\sigma}\tilde R^{\mu\nu\rho\sigma},
\] 
where $\tilde R_{\mu\nu\rho\sigma}$ is the Riemann tensor of the
physical metric $\tilde g$. The curves are consistent with what we
have just said, and confirm in particular that the collapse or
expansion takes place \textit{globally} in space eventually.  The
kinks in these curves can be explained as follows.  The spatial
profiles of the Kretschmann scalar in our evolutions have several
local extrema in space which ``compete'' to become the global
extremum.

\begin{figure}[t]
  \centering
  \psfrag{t}[][][0.9]{$t$}
  \psfrag{Max(H)}[][][0.9]{$\text{max}(H)$}
  \psfrag{Min(Kretschmann)}[][][0.9]{$\text{min}(K)$}
  \includegraphics[width=0.49\textwidth]{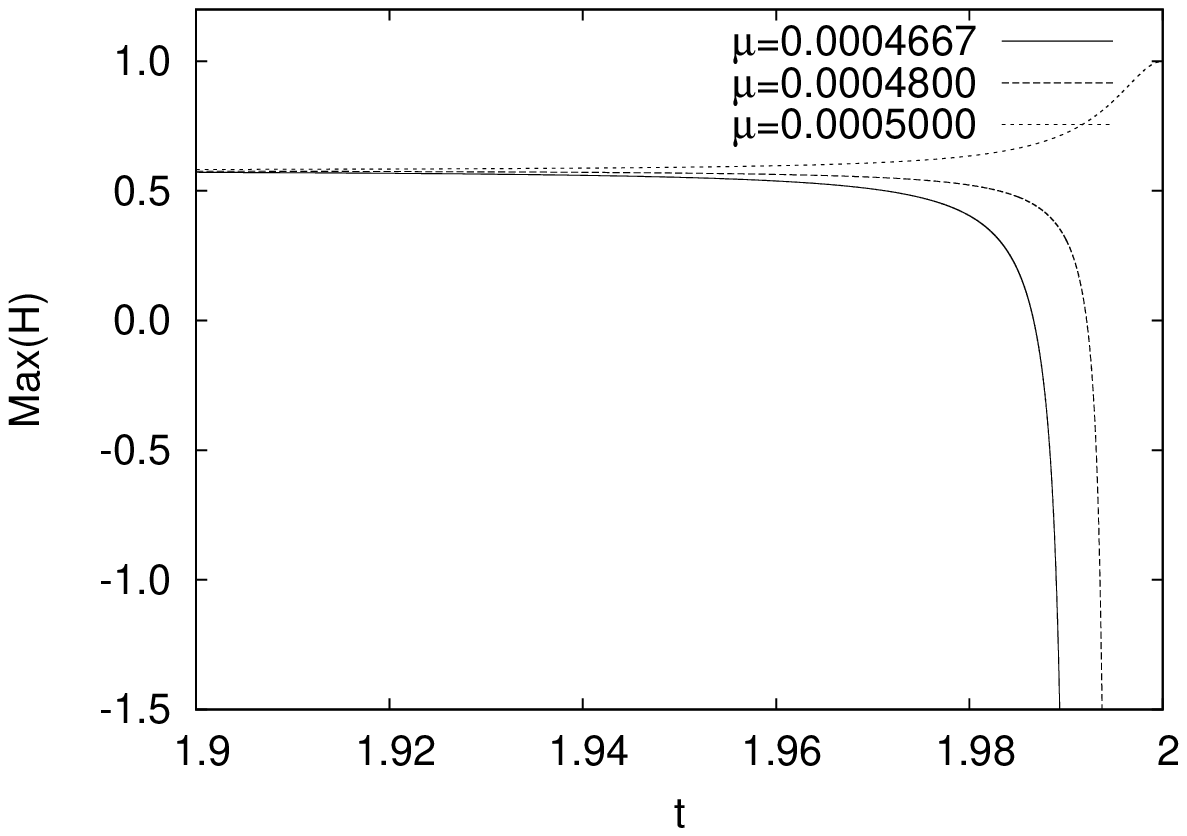}
  \includegraphics[width=0.49\textwidth]{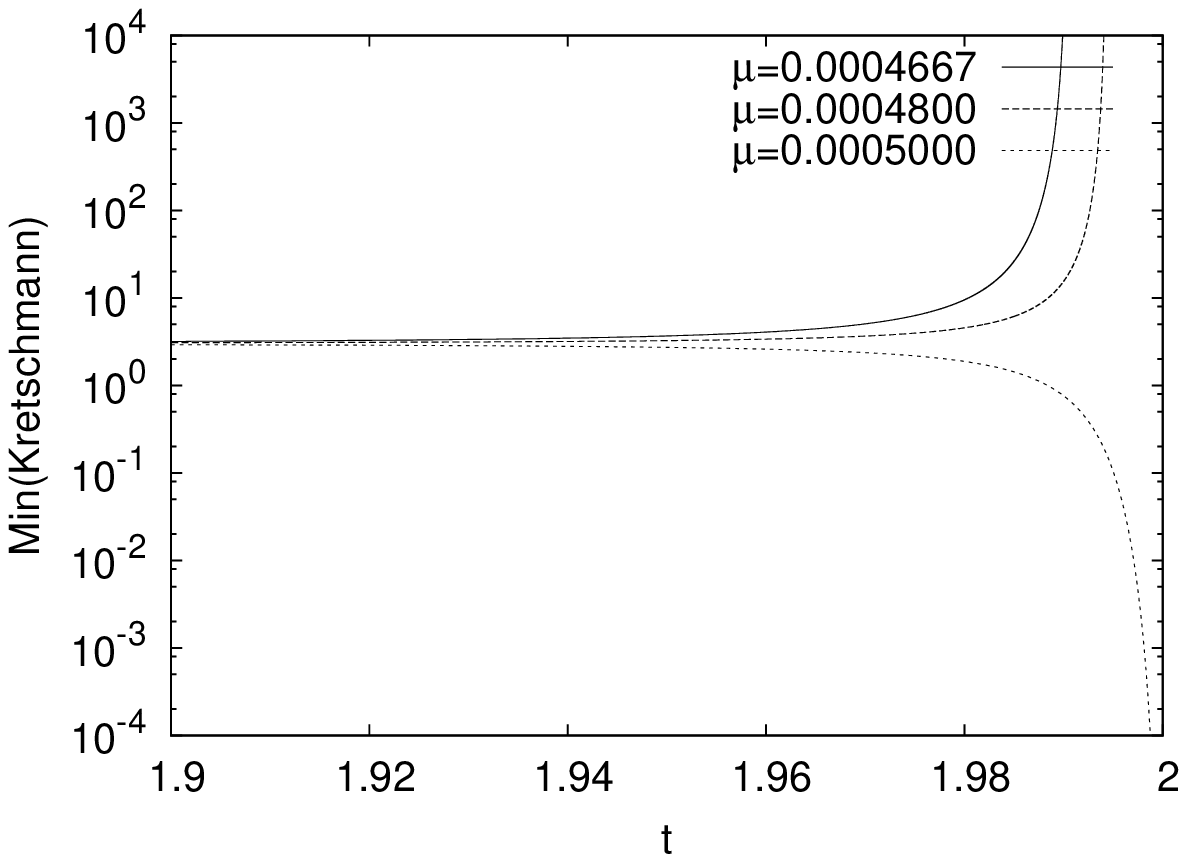}
  \caption{Future late time behavior.}
  \label{fig:latetime}
\end{figure}
All the plots so far focus on the early time behavior of the solutions
due to the choice of scales on the axes. Now let us look at
\Figref{fig:latetime}, which focuses on the evolutions at late times.
We note that on these scales, the curves of the maxima and minima of
the quantities are not distinguishable and hence we only show one. In
the first picture, we show the Hubble scalar $H$, cf.\
\Eqsref{eq:GowdyS1xS2meancurvatures}. In the eventually expanding case
given by $\mu=0.0005$, we can show that the solution develops a smooth
$\scrip$ numerically. This is consistent with (but not implied by) the
fact that $H$ converges to the value $1$ at $t=2$ in the plot. 
\opt{longversion}{%
  Recall the discussion of the behavior of $H$ at $\scrip$
  in \Sectionref{P1-sec:cosmicnohairnariai} of the first paper
  \cite{beyer09:Nariai1} for the case $\Lambda=3$.  
}%
\opt{shortversion}{%
  Recall that $\Lambda=3$.
}%
For the other two solutions, these plots confirm that they collapse
indefinitely. This follows from a singularity theorem
\cite{galloway2002}, because in both cases, $H$ eventually becomes
smaller than $-1$.  The second plot in \Figref{fig:latetime} shows $K$
versus $t$, and it reinforces our previous statement that the
curvature of the ``collapsing'' solutions blows up everywhere in space
eventually.

\begin{figure}[t]
  \centering
  \psfrag{t}[][][0.9]{$t$}
  \psfrag{Max(H1)}[][][0.9]{$\text{max}(H_1)$}
  \includegraphics[width=0.49\textwidth]{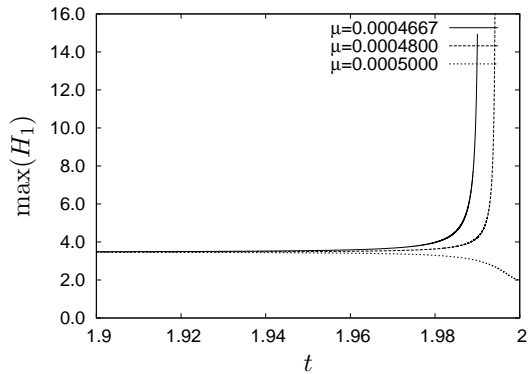}
  \caption{Future evolution of the expansion of \So-factor.}
  \label{fig:expansionS1}
\end{figure}
\begin{figure}[t]
  \centering
  \psfrag{t}[][][0.9]{$t$}  
  \psfrag{Max(sinalpha)}[][][0.9]{$\text{max}(\sin\alpha)$}
  \psfrag{Min(sinalpha)}[][][0.9]{$\text{min}(\sin\alpha)$}  
  \includegraphics[width=0.49\textwidth]{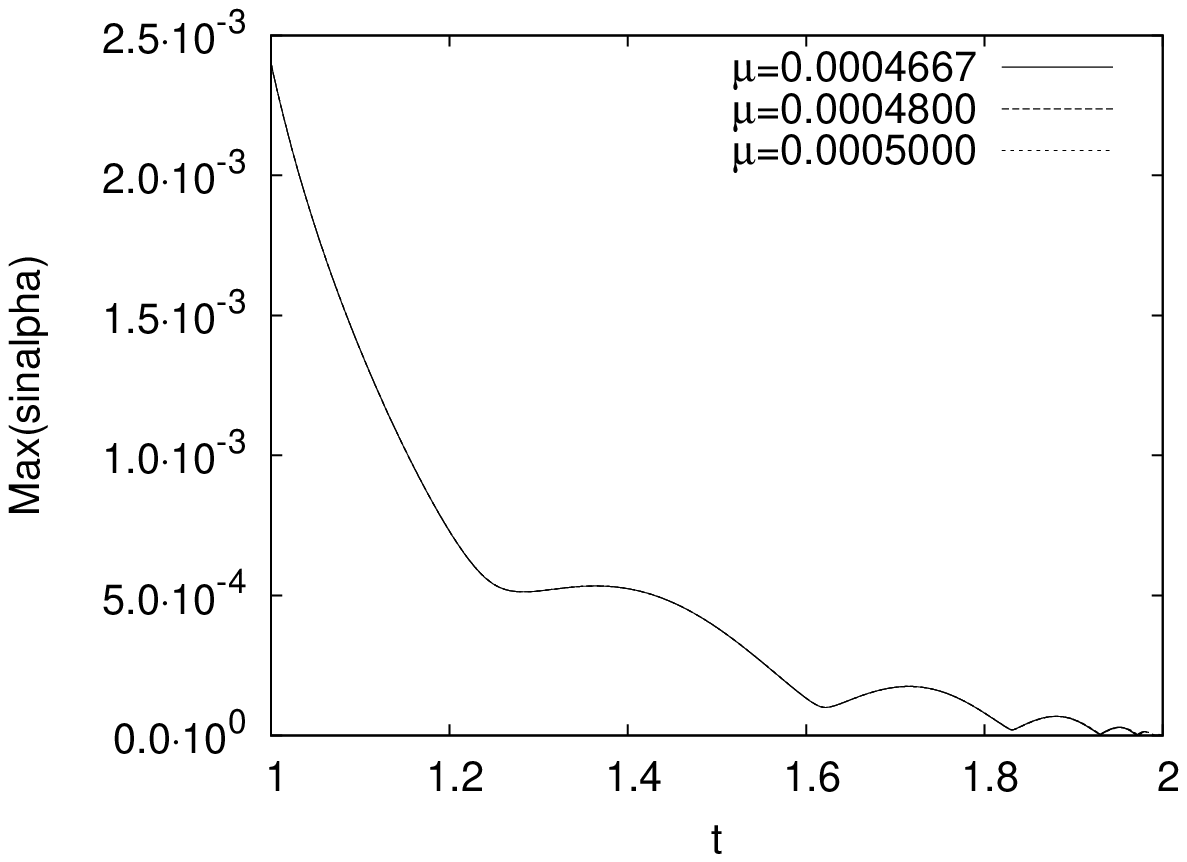}
  \includegraphics[width=0.49\textwidth]{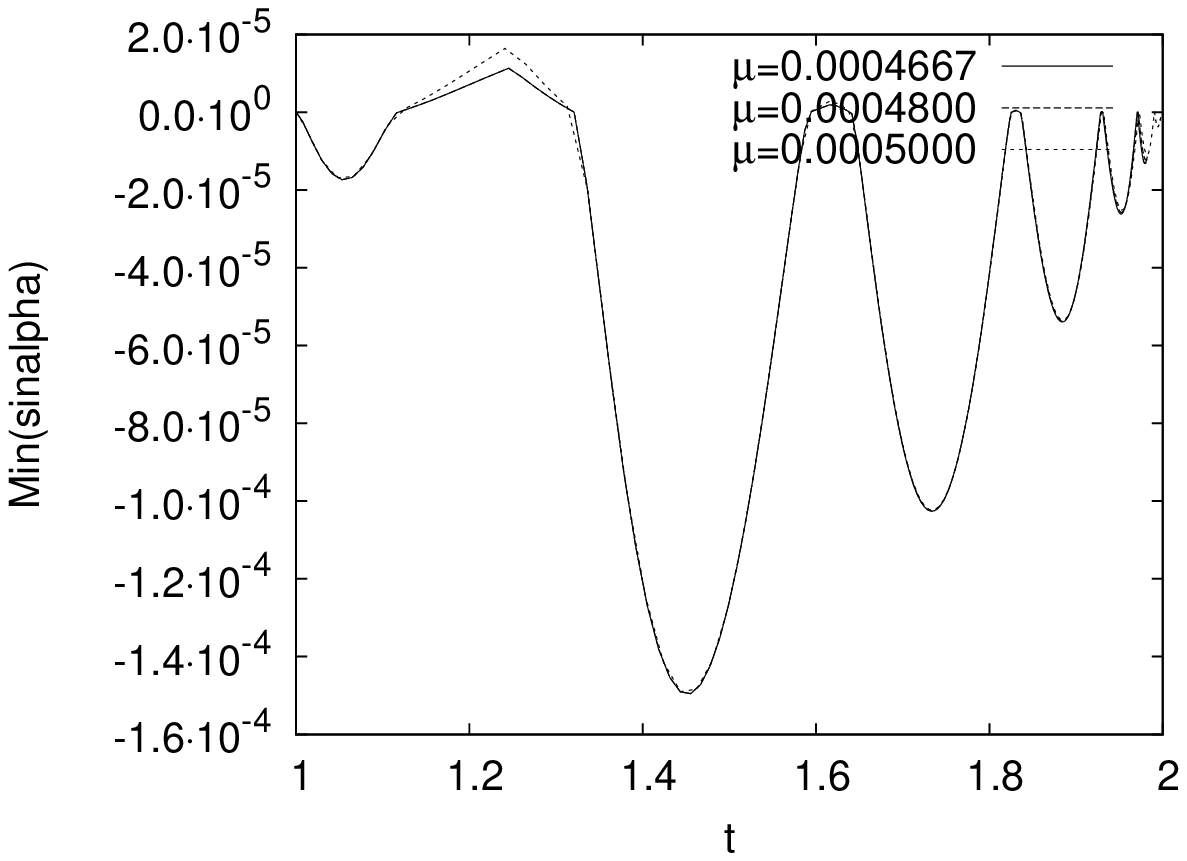}
  \caption{Future evolution of the angle between the \So- and the
    normal of the \St-factor.}
  \label{fig:angleS1S2}
\end{figure}
Let us finish this part with a discussion of the evolution of other
aspects of the geometry.  In \Figref{fig:expansionS1}, we see the
evolution of $H_1$, cf.\ \Eqsref{eq:GowdyS1xS2meancurvatures}, i.e.\
the expansion of the coordinate $\So$-factor. According to this plot,
we conjecture that the two collapsing solutions form a singularity of
cigar type \cite{Wainwright}, in the same way as in the spatially
homogeneous case \cite{beyer09:Nariai1}. In \Figref{fig:angleS1S2}, we
show the maximum and minimum value of $\sin\alpha$, where the angle
$\alpha$ is defined as follows. At a given time $t=const$ and spatial
point, $\alpha$ is the angle between the vector $\partial_\rho$ and
the normal vector of the coordinate $\St$-factor within the
$t=const$-surface. The plots suggest that $\alpha$ approaches zero
eventually. Hence, loosely speaking, the Gowdy solutions become more
and more polarized. Again, we do not understand the mechanisms
underlying these curves and hope that a linearization will shed
further light on this. Since the curves for the three solutions are
almost indistinguishable, it is a natural question whether this
behavior is universal in our class of solutions.

\paragraph{Past evolution}
\begin{figure}[t]
  \centering
  \psfrag{t}[][][0.9]{$t$}
  \psfrag{Max(H)}[][][0.9]{$\text{max}(H)$}
  \psfrag{Max(Kretschmann)}[][][0.9]{$\text{max}(K)$}
  \psfrag{Min(Kretschmann)}[][][0.9]{$\text{min}(K)$}
  \includegraphics[width=0.49\textwidth]{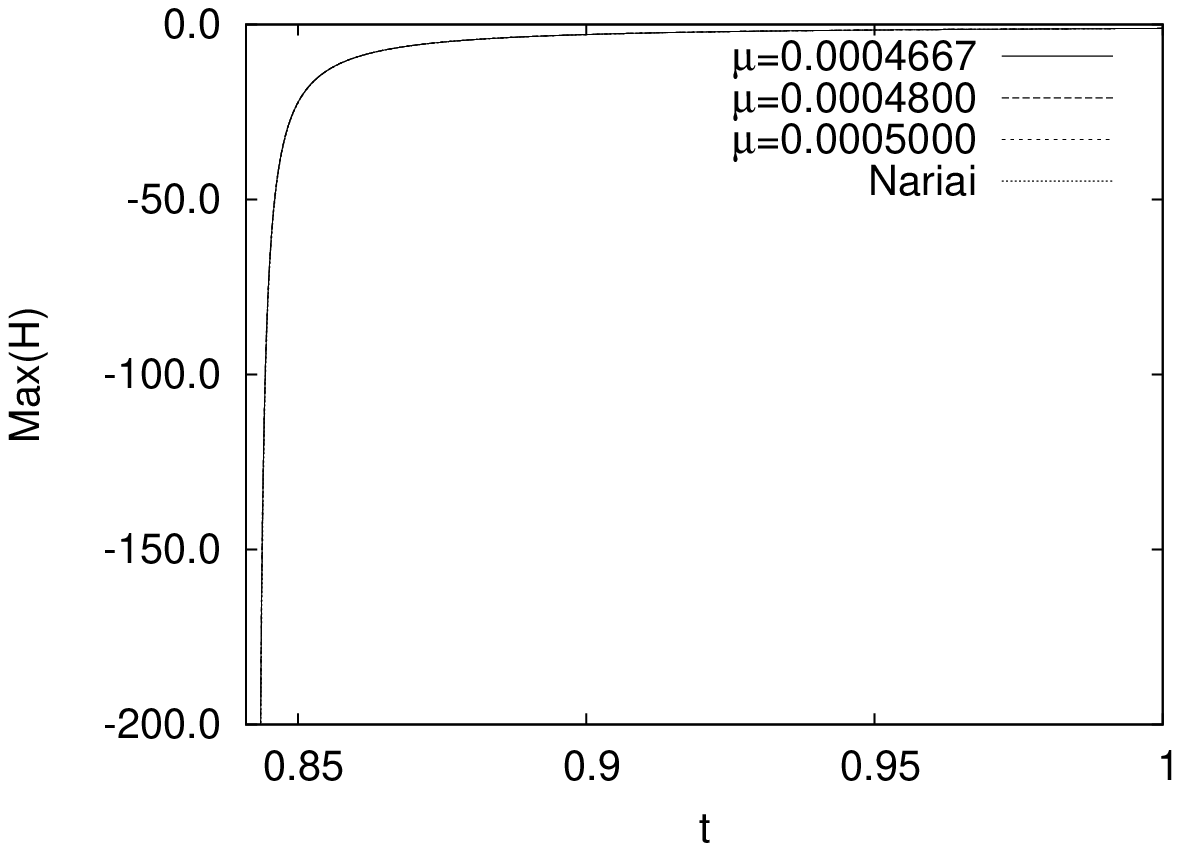}\\
  \includegraphics[width=0.49\textwidth]{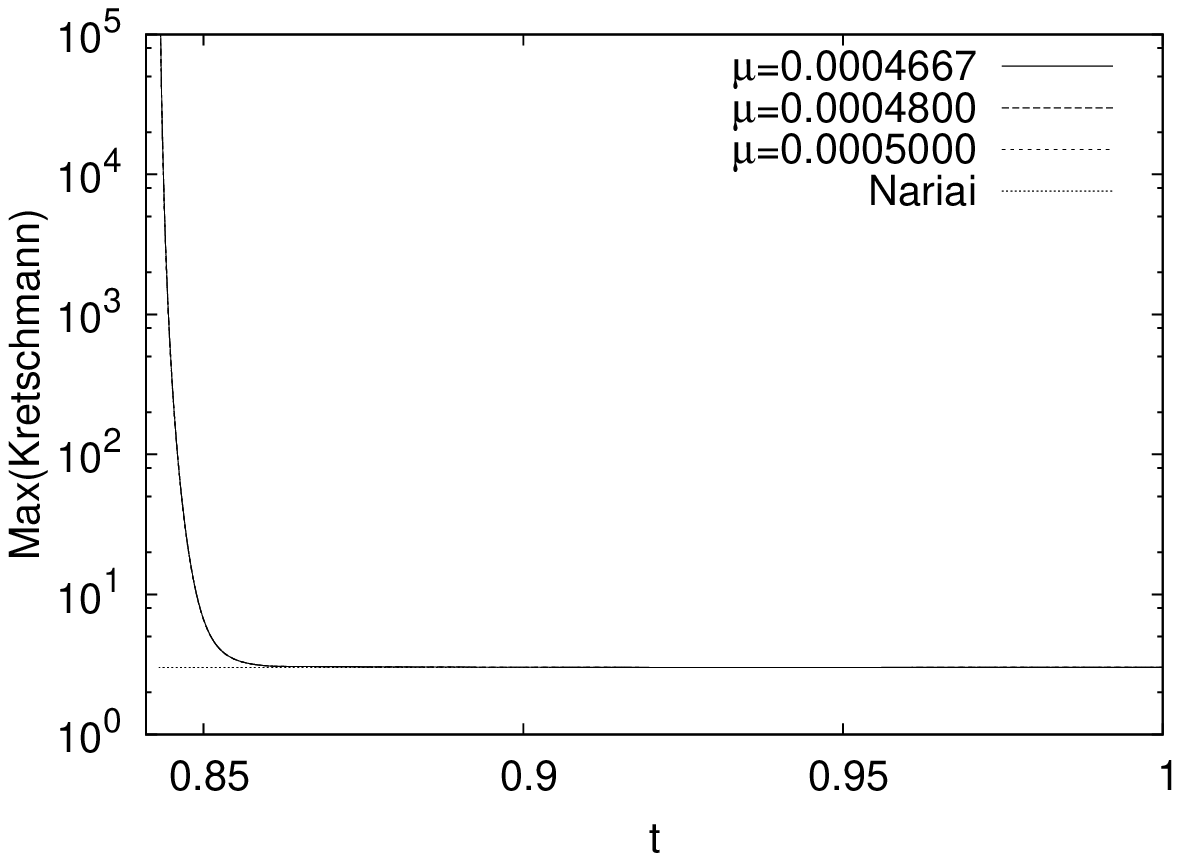}
  \includegraphics[width=0.49\textwidth]{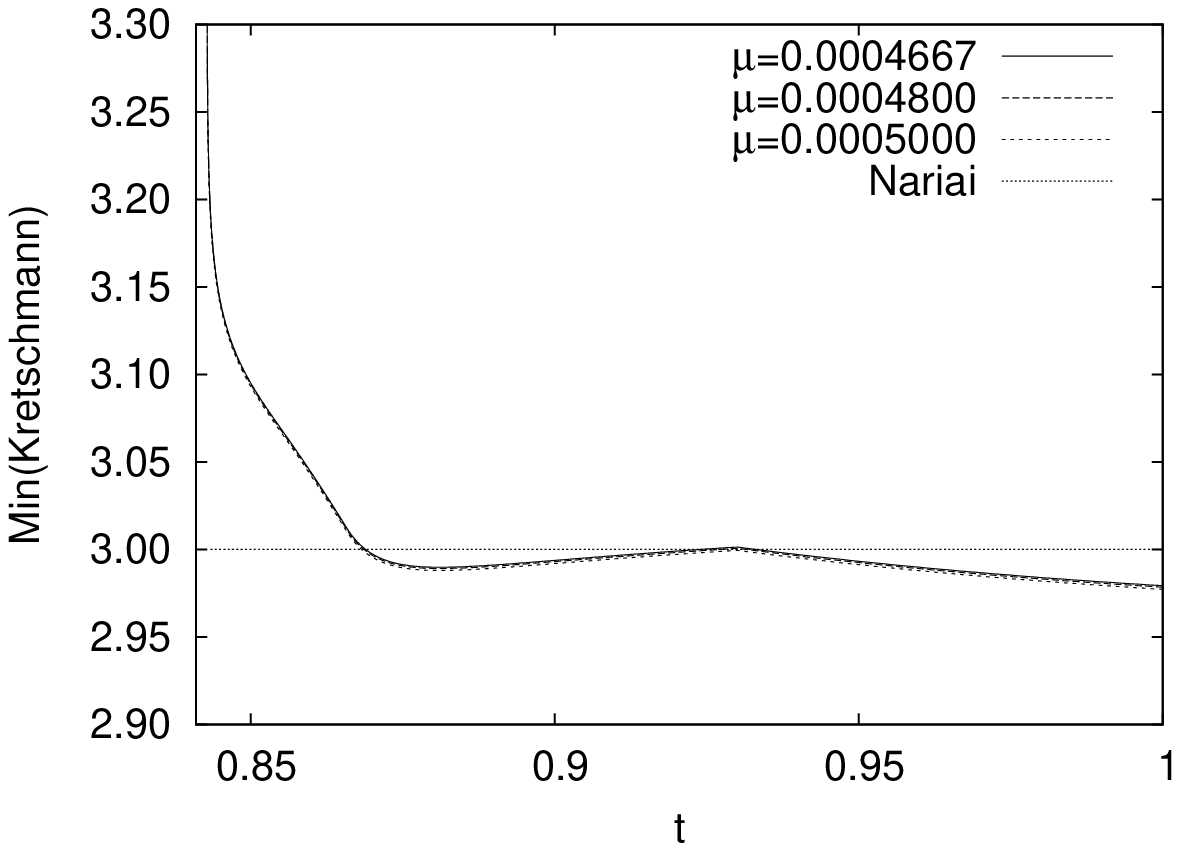}
  \caption{Past evolution.}
  \label{fig:pastevolution}
\end{figure}
For completeness, let us proceed with the evolution in the past time
direction, and recall from the discussion in the first paper
\cite{beyer09:Nariai1} that the unperturbed Nariai solutions with
$\sigma_0<0$ form a Cauchy horizon in the past. There is additional
motivation from the strong cosmic censorship issue \cite{andersson04a}
to understand what happens to this horizon under our perturbations.
Because we consider the past time direction now, the initial
hypersurface given by $t=1$ is on the right of the following plots,
and the past evolutions take place to the left. In the first plot of
\Figref{fig:pastevolution}, we show the maximum of the Hubble scalar
$H$ for our three cases of initial data together with the
corresponding curve of the unperturbed Nariai solution. Again,
corresponding curves of the minima are not distinguishable on these
scales. We see that the four curves in the plot are almost the same,
and hence all four cases collapse in the same way to the past. Do the
perturbed solutions hence also develop a Cauchy horizon in the past?
The second and third plot in \Figref{fig:pastevolution} suggest that
this is not the case, as the curvature blows up uniformly for all
perturbed solutions. We stress that the numerical results for the
minimum of the Kretschmann scalar are not conclusive yet, since the
value of $\text{min}(K)$ is still relatively small when the runs were
stopped. Nevertheless, first signs of curvature blow up are apparent.
Our observations in \cite{beyer08:TaubNUT} were quite similar, and
$\text{min}(K)$ often blew up much less than $\text{max}(K)$ close to
a singularity. It is expected that this is not a geometrical
phenomenon, but rather caused by the choice of the Gauss gauge, as we
discuss there.

\subsection{Practical details about the runs and numerical errors}
\label{sec:DetailsErrors}

\paragraph{Further technical details}
\begin{table}[t]
  \centering
  \begin{tabular}{|cc|ccc|ccccc|}\hline
    direction & $\mu$ & $N_0$ & $N_1$ & $\mu$ & $h_0$ & $h_1$ & $h_{min}$ & $\eta$ & $t_1$\\\hline
    future &  $0.0004667$ & $300$ & $300$ & $10^{-10}$ & $10^{-3}$ & $10^{-7}$ & $10^{-7}$ & $10^{-15}$ & $1.990$\\
    future &  $0.0004800$ & $300$ & $300$ & $10^{-10}$ & $10^{-3}$ & $10^{-7}$ & $10^{-7}$& $10^{-15}$ & $1.994$\\
    future &  $0.0005000$ & $200$ & $200$ & $10^{-10}$ & $10^{-3}$ & $10^{-6}$ & $10^{-6}$& $10^{-14}$ & $2.000$\\
    past &  $0.0004667$ & $200$ & $300$ & $10^{-11}$ & $10^{-3}$ & $10^{-6}$ & $10^{-6}$& $10^{-14}$ & $0.842$\\
    past &  $0.0004800$ & $200$ & $300$ & $10^{-11}$ & $10^{-3}$ & $10^{-6}$ & $10^{-6}$& $10^{-14}$ & $0.842$\\
    past &  $0.0005000$ & $200$ & $300$ & $10^{-11}$ & $10^{-3}$ & $10^{-6}$ & $10^{-6}$& $10^{-14}$ & $0.842$\\\hline
  \end{tabular}  
  \caption{Numerical parameters for the runs presented in 
    \Sectionref{sec:thenumresults}.}
  \label{tab:numericalparameters}
\end{table}
Our general numerical setup has been described
in \Sectionref{sec:numerics}. In Table~\ref{tab:numericalparameters}
now, we list more technical details about the runs in the previous
section. The quantities $N_0$, $N_1$ and $\mu$ are related to the
spatial resolution. Our numerical runs use the simple spatial adaption
technique described in \cite{beyer08:code}. After some experiments,
the quantity $\chi_{22}$ was chosen as the reference variable. The
threshold value for the spatial adaption is called $\mu$, the initial
number of spatial grid points is $N_0$, and the number of spatial grid
points at the stop time $t_1$ is referred to as $N_1$. The following
columns in the table describe the time discretization. We use the
$5$th-order ``embedded'' adaptive Runge Kutta scheme with control
parameter $\eta$. This parameter was introduced in \cite{beyer08:code}
in order to control the desired accuracy of the time integration; the
lower its value is, the smaller are the time steps chosen by the
adaption algorithm.

Furthermore, $h_0$ is the initial time
step and $h_1$ is the time step at the stop time $t_1$. In order to
prevent the code from reaching unpractically small values of $h$, the
adaption is switched off when $h$ goes below $h_{min}$. One sees that
for all the runs, this minimum value was reached eventually. Note that
all numbers in the table are rounded.

\paragraph{Numerical errors and convergence}
Prior to the numerical runs in the previous section, we made
further tests of the code in addition to those in \cite{beyer08:code}. The
choice of orthonormal frame in \Sectionref{sec:numcompID} has the
consequence that even spatially homogeneous solutions ``appear
inhomogeneous'', in the sense that many resulting unknown tensor
components depend on the spatial coordinates. Spatially homogeneous
solutions hence yield a non-trivial test case for the code. These
tests showed that the code is able to reproduce these solutions with
promisingly small errors, in particular the Nariai solution itself.

\begin{figure}[t]
  \psfrag{t}[][][0.9]{$t$}
  \psfrag{|E11(0) Difference|}[][][0.9]
  {Convergence $e\indices{_1^1}(t,0)$}
  \begin{minipage}{0.49\linewidth}
    \centering 
    \includegraphics[width=\textwidth]{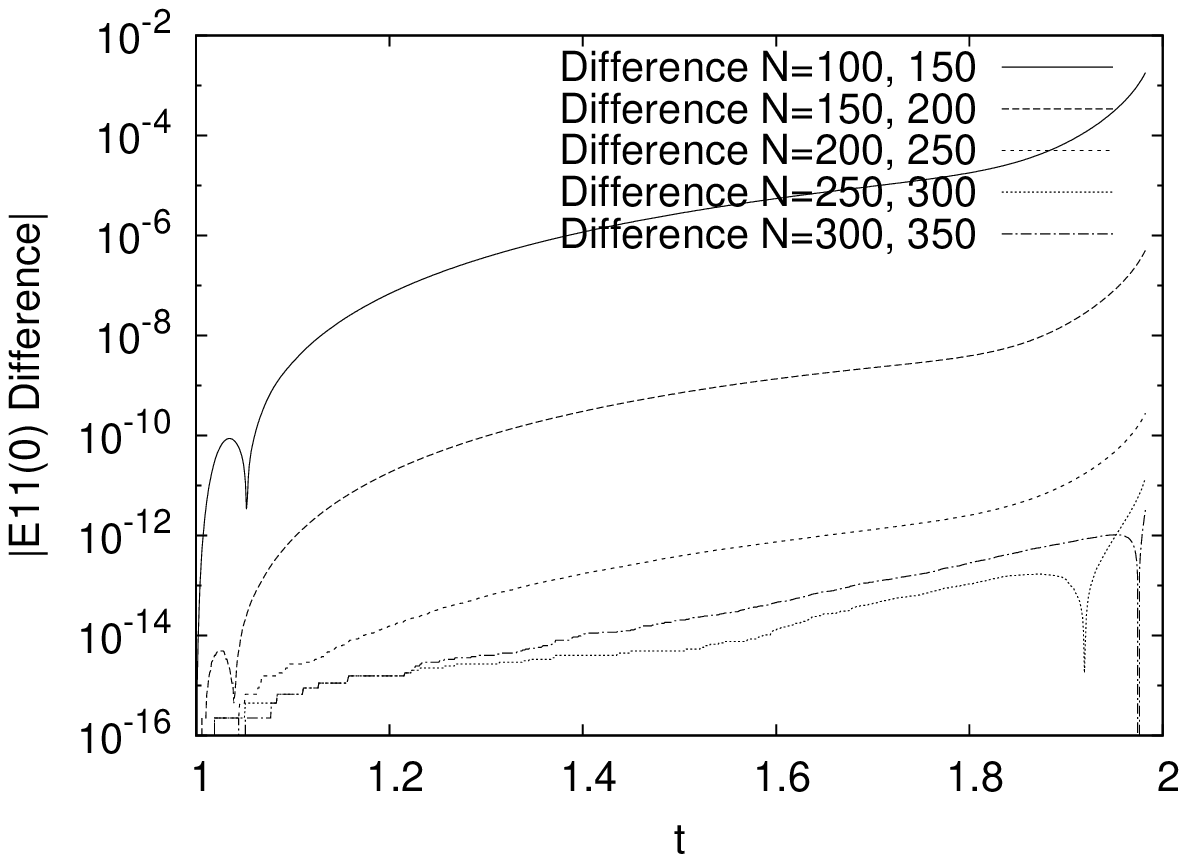}
    \caption{Spatial convergence for $\mu=0.00048$.}
    \label{fig:spat_conv}
  \end{minipage}
  \begin{minipage}{0.49\linewidth}
    \centering 
    \includegraphics[width=\textwidth]{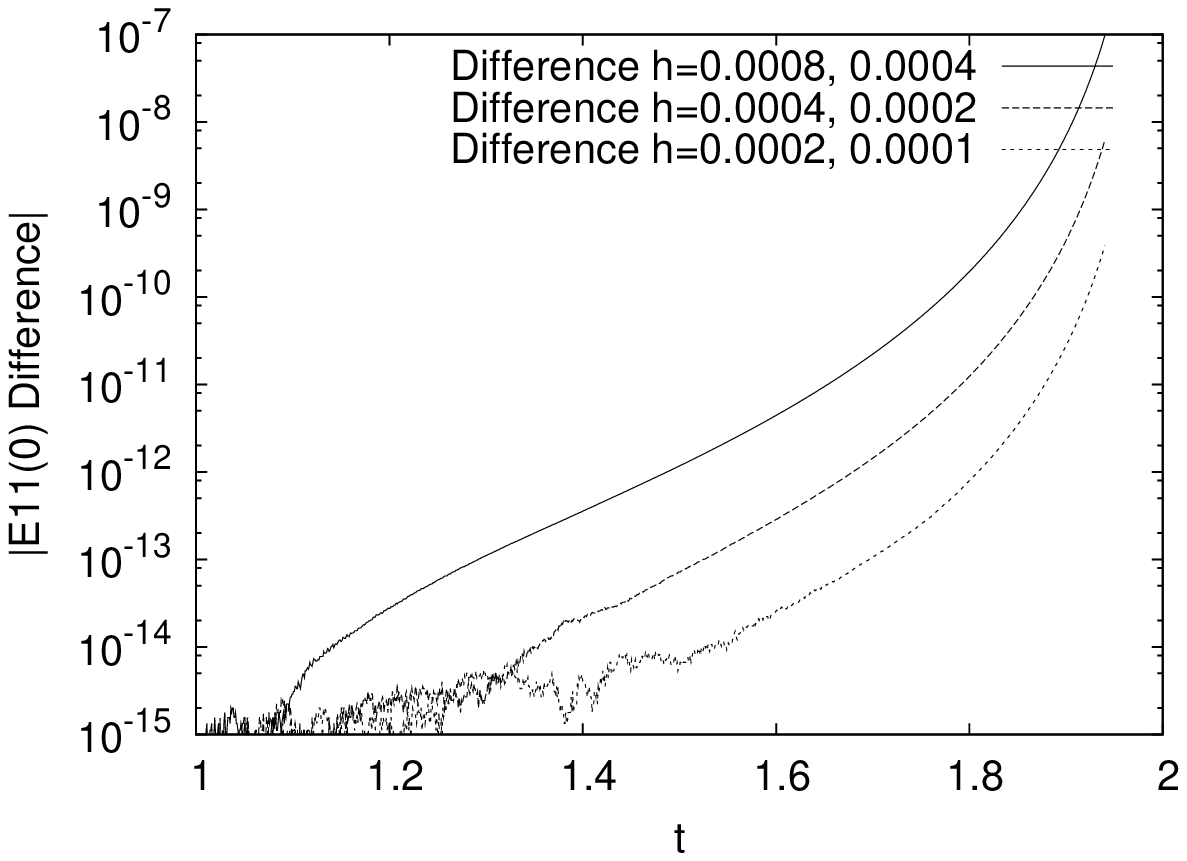}
    \caption{Time convergence for $\mu=0.00048$.}
    \label{fig:time_conv}
  \end{minipage}
\end{figure}
In order to give the reader an impression of the size of numerical
errors in the results in the previous section, let us redo the run for
the initial data set $\mu=0.00048$ and $\Sigma_\times^{(1)}=4\cdot
10^{-4}$ to the future with other resolutions than in
Table~\ref{tab:numericalparameters}. In order to study convergence
more cleanly, let us switch off all adaption techniques for
this. First, consider \Figref{fig:spat_conv}. For the same initial
data, we made six runs with the spatial resolutions $N=100$, $150$,
$200$, $250$, $300$ and $350$, and fixed size of time step
$h=10^{-4}$. The figure shows the absolute values of the differences
of two successive runs for the quantity $e\indices{_1^1}$ at
$\theta=0$ versus time $t$. We note that we have looked at other
variables and seen the same results qualitatively. The absolute size
of these differences can be interpreted as a measure for the size of
the absolute pure numerical error for a given spatial resolution $N$
(or equivalently spectral truncation) on the one hand. On the other
hand, since these errors get smaller dramatically for increasing $N$,
we have demonstrated convergence of the error in our numerical
results.  Up to $N\approx 300$, the numerical errors in these runs are
hence dominated by the spectral discretization. Increasing $N$
further, does not decrease the numerical error, and other types of
errors become dominant, in particular the errors given by the time
discretization and machine round-off errors. The plot also allows us to
quantify the rate of convergence. For resolutions smaller than
$N\approx 250$, we find that $50$ additional grid points decrease the
error by a factor of approximately $3000$. This shows that the
convergence is exponential in this regime, and hence confirms that our
numerical techniques are reliable and the numerical errors in our
results in \Sectionref{sec:thenumresults} are small.  The fact that
spatial resolutions $N\approx 300$ are necessary in order to make
spatial discretization errors smaller than other errors even at early
times, when the solutions are very smooth in space in principle, shows
that our choice of frame is not optimal. However, the fact that we see
such a nice convergence for a quantity evaluated at the coordinate
singularity $\theta=0$ provides particular evidence that our numerical
regularization of the coordinate singularities mentioned
in \Sectionref{sec:numerics} works well.

In \Figref{fig:time_conv}, we show the same for \textit{fixed} spatial
resolution $N=300$ and the following time resolutions: $h=8\cdot
10^{-4}$, $4\cdot 10^{-4}$, $2\cdot 10^{-4}$ and $1\cdot 10^{-4}$. As
mentioned earlier, the errors given by the spatial discretization
should be negligible for $N=300$. Note, that here, instead of the
adaptive $5$th order Runge Kutta scheme, we use the standard $4$th
order Runge Kutta scheme. The figure confirms $4$th order convergence
of the errors as long as those are dominated by time
discretization. This is the case in particular for later evolution
times. At very early times, however, the errors are strongly
influenced by the machine round-off errors and so barely converge with
increasing resolution. If we decreased $h$ even further, the errors
would be more and more dominated by round-off errors for longer and
longer evolution times and convergence would be lost. Again, all this
confirms that our numerical techniques are reliable and the numerical
errors in our results in \Sectionref{sec:thenumresults} are well
understood and small.

For the discussion of other important error quantities, we introduce
the following definitions from \cite{beyer08:code,beyer08:TaubNUT}.
First, we define
\[\normeinstein(t)
:=\left\|(\tilde R_{ij} -\lambda \tilde
  g_{ij})/\Omega\right\|_{L^{1}(\Sigma_t)},
\] 
with the physical Ricci tensor $\tilde R_{ij}$ evaluated algebraically
from the conformal Schouten tensor $L_{ij}$ and derivatives of the
conformal factor $\Omega$. The spatial slice at time $t$ is referred
to as $\Sigma_t$ here. The indices involved in this expression are
defined with respect to the physical orthonormal frame given by
$\tilde e_i=\Omega e_i$, and we sum over the $L^1$-norms of each
component. Hence, this norm yields a measure of how well the numerical
solution satisfies Einstein's field equations \Eqref{eq:EFE}. Second,
let us define $\normconstr$ as the $L^1$-norm of the sum of the
absolute values of each of the six components of the left hand sides
of \Eqsref{eq:bianchi_constraints} at a given instant of time $t$. For
the definition of the norm $\normbc$, we again refer to
\cite{beyer08:code}. The smoothness of the solution implies a certain
behavior of all unknowns in our evolution problem at the coordinate
singularities at $\theta=0,\pi$, and the quantity $\normbc$ is the sum
of the absolute values of all quantities, that, in line with this
behavior, should vanish at $\theta=0,\pi$ at a given time of the
evolution.

\begin{figure}[t]
  \psfrag{t}[][][0.9]{$t$}
  \psfrag{NormConstr}[][][0.9]{$\normconstr$}
  \psfrag{NormEinstein}[][][0.9]{$\normeinstein$}
  \begin{minipage}{0.49\linewidth}
    \centering
    \includegraphics[width=\textwidth]{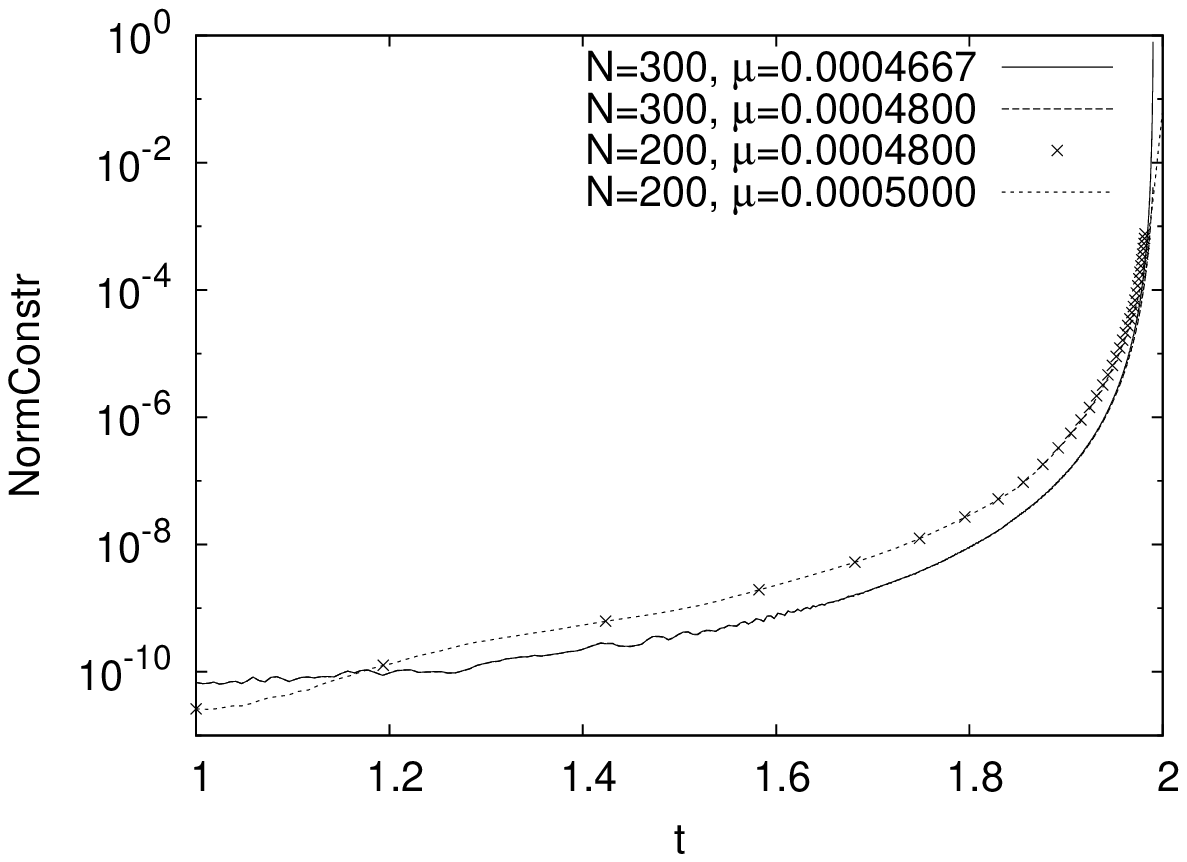}
    \caption{Violation of the constraints.}
    \label{fig:constrviol}
  \end{minipage}
  \begin{minipage}{0.49\linewidth}
    \centering
    \includegraphics[width=\textwidth]{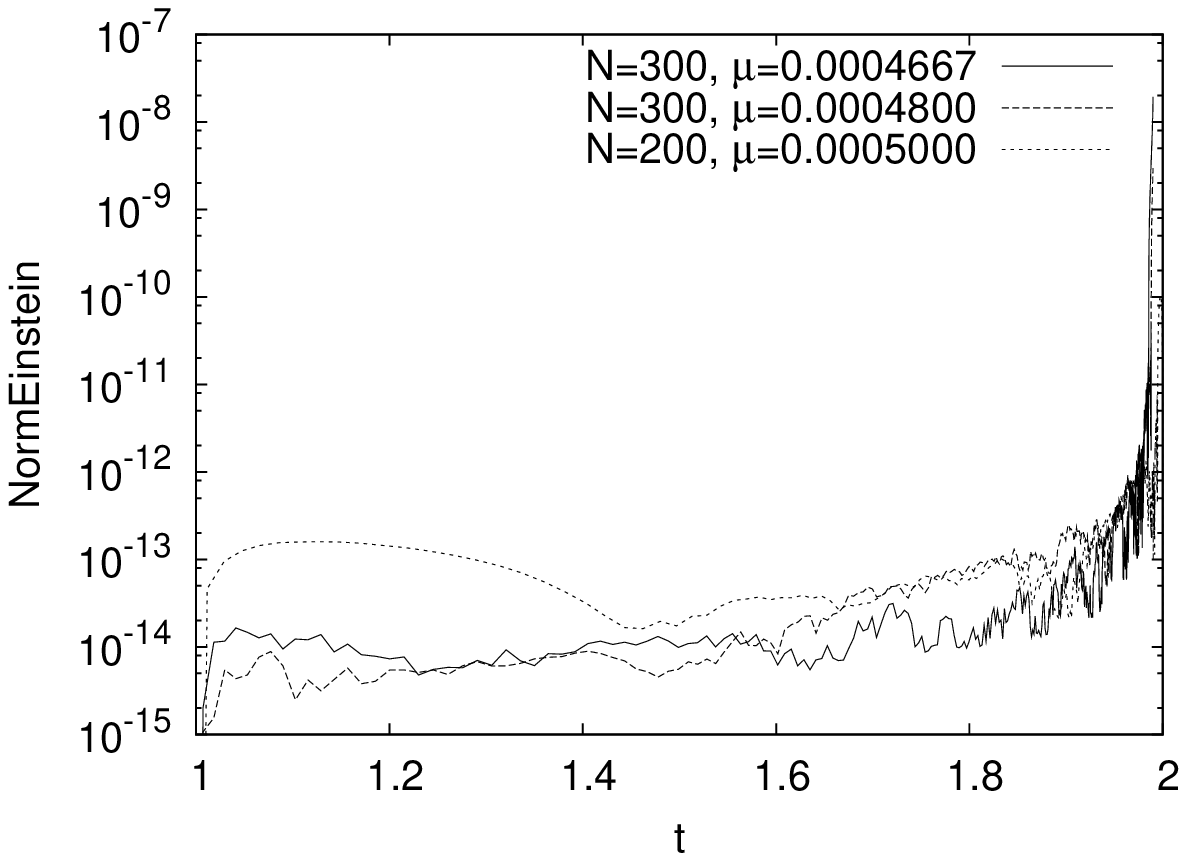}
    \caption{Violation of Einstein's field eqs.}
    \label{fig:EFEviol}
  \end{minipage}
\end{figure}  
These norms are used in the following. In addition to the pure
numerical errors of the type discussed above, numerical relativity is
plagued with the ``continuum instability'' of the constraint
hypersurface in general when the constraints are propagated freely. We
hence stress that this is not a particular problem of our numerical
investigations here.  For some evolution systems, one is able to
control the constraint behavior slightly \cite{Gundlach05}, but a
general solution to this fundamental problem has not yet been
found. When the evolution is started with an arbitrary small violation
of the constraints, then typically, these violations grow
exponentially, or even blow up after finite time, even for the
continuum (i.e.\ non-discretized) equations \cite{Friedrich05}. In
\Figref{fig:constrviol}, we see the constraint violations for the runs
to the future of the previous section.  In addition, we show the
constraint propagation for the case $\mu=0.00048$ and
$\Sigma_\times^{(1)}=4\cdot 10^{-4}$ with the same numerical
parameters as for the case $\mu=0.0005$ from
Table~\ref{tab:numericalparameters}. In accordance with typical
numerical runs, we see that the constraint violations grow strongly
during the evolution, almost in the same way whether the solutions
collapse or expand eventually. Increasing the initial spatial
resolution leads to higher initial constraint violations due to
initially higher machine round-off errors.  Nonetheless, the plot
suggests that the constraint violations decrease once the
discretization errors become dominant.  This positive result is
consistent with our observations in \cite{beyer08:TaubNUT}, and
demonstrates that the constraint violations can be controlled and kept
close to the continuum evolution of the constraints for arbitrary
large evolution times as long as the errors are dominated by
discretization and not by machine round-off errors.  As discussed
above, for the continuum equations, the initial value of the
constraint violation is of fundamental importance. We have tested this
for $N=300$ in the following way, as we do not show here. We have
repeated some of the runs before with ``quad precision'', mentioned
in \Sectionref{sec:numerics}. With quad precision, the initial size of
the constraint violation is many orders of magnitude smaller than in
the standard ``double precision'' case. Our numerical evolutions show
that this stays true for the whole evolution in particular because the
machine round-off errors, which would otherwise dominate for $N=300$
and sufficiently high time resolution, are much smaller.  Now, since
the results presented in \Sectionref{sec:thenumresults} are virtually
unchanged when they are repeated with quad precision, we conclude that
these results are reliable despite the apparently large violation of
the constraints at late times. Concerning the violation of the full
Einstein's field equations, hence including all constraints and
evolution equations, the same arguments lead to similar
conclusions; consider \Figref{fig:EFEviol}.

\begin{figure}[t]
  \psfrag{t}[][][0.9]{$t$}
  \psfrag{ViolBC}[][][0.9]{$\normbc$}  
    \centering  
    \includegraphics[width=0.49\textwidth]{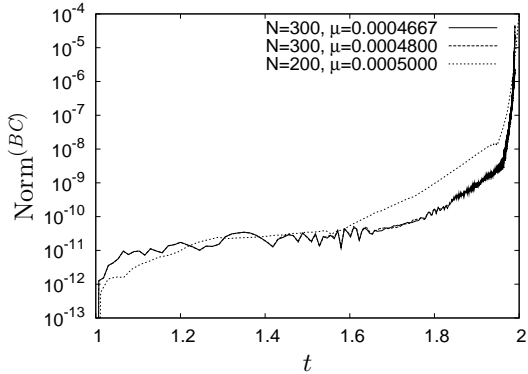}
    \caption{Violation of boundary conditions.}
    \label{fig:violBC}
\end{figure}
Let us point the attention of the reader to \Figref{fig:violBC}, in
order to show the order of magnitude of the violation of the
smoothness conditions at the coordinate singularities.  As we do not
show here, these errors converge to zero, as long as the pure
numerical errors are not dominated by machine round-off errors. We
note that none of the runs presented here enforce these smoothness
conditions explicitly; it is possible that this would improve the
numerical accuracy slightly \cite{beyer08:code}.


%% file: summary.tex
In this paper, we studied the instability (non-genericity) of the
Nariai solutions for the family of Gowdy perturbations. The
investigations here are based on the first paper
\cite{beyer09:Nariai1}. Our motivations to do this were
two-fold. First, we were interested in the fundamental question of
cosmic no-hair and its dynamical realization in more general classes
than the spatially homogeneous case considered in
\cite{beyer09:Nariai1}. Second, the results of the first paper suggest
that the understanding of the instability of the Nariai solutions in
the spatially homogeneous case could be exploited in order to
construct cosmological black hole solutions with in principle
arbitrarily complicated combinations of black hole and cosmological
regions. Indeed, this interesting possibility was already considered
in the spherically symmetric case in \cite{Bousso03}, where the author
claims that such constructions are possible. Since no non-trivial
cosmological black hole solutions are known for Gowdy symmetry with
spatial \SoXSt-topology to our knowledge, it was our aim to address
this open problem. 

Our results, which are obtained with the numerical technique
introduced in \cite{beyer08:code}, are as follows. First, by making
experiments with various choices of perturbations, indeed more than
those presented in this paper, we can confirm the expected instability
of Nariai solutions, and hence the cosmic no-hair conjecture also in
the case of Gowdy symmetric perturbations of the Nariai solution. That
is, either the solutions close to a Nariai spacetime collapse in a
given time direction, or when they expand, they form a smooth
conformal boundary and hence are consistent with the cosmic no-hair
picture. Hence, our results can be seen as a generalization of the
work in \cite{beyer09:Nariai1} on the one hand. However, of even
stronger interest is that our numerical results suggest that it is
\textit{not} possible to construct cosmological black hole solutions
with small Gowdy symmetric perturbations of the Nariai solutions. This
result is unexpected, in particular it is contrary to the claims for
spherical symmetry. We find that the early time behavior agrees with
the expectations. But then, the quantity $H_2$ starts to level off and
the solution makes a decision, whether to either expand or collapse
\textit{globally} in space. The underlying mechanism is not
understood. Of particular interest for future research will be the
construction and study of the critical solution and of possible
critical phenomena. One promising approach for shedding further light
on these issues is to linearize the problem, on the one hand around
the unperturbed Nariai solution, and on the other hand around the
hypothetical critical solution.

Certainly, our class of initial data cannot be considered as
``generic'', or to put it the other way around, it is not clear how
``special'' it is. Thus it is hard to make predictions for general
solutions close to generalized Nariai spacetimes. We are currently
working on a method to obtain ``general'' Gowdy symmetric initial data
numerically. General Gowdy initial data would allow us to study
generalized Nariai solutions in particular in the standard case
$\sigma_0> 0$. In this light, we understand our results here as first
steps in an ongoing research project.

Since we find that it does not seem to be possible to construct
cosmological black hole solutions by means of small Gowdy symmetric
perturbations of Nariai data, it is natural to investigate large
perturbations as a next step.  The hope is that the spatially local
behavior, which is suppressed in the case of small perturbations
apparently, becomes significant. Beyond what we have presented in this
paper here, our preliminary results suggest that this is the
case. This will be investigated in another future publication.

We have discussed numerical errors and given some evidence that our
numerical results are reliable. However, there is certainly room for
improvements, not only in the numerical techniques, but also in the
choice of gauge and the particular formulation of the field equations.
For instance, in \cite{beyer08:TaubNUT}, we have interpreted the fact
that we do not see spatially local behavior close to the singularities
in our runs, so-called Gowdy spikes \cite{andersson04a}, as a
reflection of the ``bad'' features of the Gauss gauge. That is, in
this gauge, the solution approaches the singularity in a too
inhomogeneous manner, obscuring such small scale structure. Hence,
other gauge choices should be investigated. Another problem, already
addressed before, is that our particular evolution system does not
show optimal constraint propagation. Other formulations of the system
should be tried and ``constraint damping terms'' should be
investigated in order to improve this problem. Nevertheless, we have
concluded above that our current numerical results can be trusted
despite the apparently large constraint violations.


%% file: acknowledgements.tex
This work was supported in part by the Göran Gustafsson Foundation,
and in part by the Agence Nationale de la Recherche (ANR) through the
Grant 06-2-134423 entitled Mathematical Methods in General Relativity
(MATH-GR) at the Laboratoire J.-L. Lions (Universit\'e Pierre et Marie
Curie).  Some of the work was done during the program ``Geometry,
Analysis, and General Relativity'' at the Mittag-Leffler institute in
Stockholm in fall 2008. I would like to thank in particular Helmut
Friedrich and Hans Ringström for helpful discussions and explanations.


%% file: gowdysymmetry_appendix.tex
\section{Relation of the 
  \texorpdfstring{$\U\times\U$}{U(1)xU(1)}-actions on
  \texorpdfstring{\SoXSt}{S1xS2} and \texorpdfstring{\Sth}{S3}}
\label{sec:actionU1xU1S1xS2S3}

Let us consider a smooth global effective action of the group
$\U\times\U$ on a $3$-dimensional manifold. One can show, see the
references in \cite{chrusciel1990}, that the only compatible smooth
compact orientable $3$-manifolds are $\T$, $\SoXSt$, $\Sth$ and lens
spaces. Since the universal cover of the lens spaces is \Sth, they
will always be included when we speak about \Sth. In this paper, we
are particularly interested in the case \SoXSt. However, for the
discussion of our numerical approach, which was worked out for the
\Sth-case originally, it makes sense to consider the case \Sth
simultaneously now. On \SoXSt, let us consider coordinates
$(\rho,\theta,\phi)$ where $\rho\in(0,2\pi)$ is the standard parameter
on $\So$ and $(\theta,\phi)$ are standard polar coordinates on \St
\begin{equation}
  \label{eq:polarcoordinatesS2}
  x_1=\sin\theta\cos\phi,\quad
  x_2=\sin\theta\sin\phi,\quad
  x_3=\cos\theta.
\end{equation}
In writing these coordinate expressions, we assume that $\St$ is
embedded in the standard way into $\R^3$ with Cartesian coordinates
$(x_1,x_2,x_3)$. On \Sth, we consider Euler coordinates
$(\chi,\lambda_1,\lambda_2)$ with the same conventions as in
\cite{beyer08:TaubNUT}, namely
\begin{equation}
  \label{eq:eulerangleparm}
  \begin{split}
    x_1&=\cos\frac\chi 2\cos\lambda_1, 
    \quad x_2=\cos\frac\chi 2\sin\lambda_1,\\
    x_3&=\sin\frac\chi 2\cos\lambda_2, 
    \quad x_4=\sin\frac\chi 2\sin\lambda_2,
  \end{split}
\end{equation}
where we assume the standard embedding of \Sth into $\R^4$ similar to
the above. Here, $\chi\in (0,\pi)$ and $\lambda_1,\lambda_2\in
(0,2\pi)$. In terms of these coordinates on \SoXSt and \Sth,
respectively, we can write a representation of the action of the group
$G=\U\times\U$. For $\SoXSt$, one has
\[\Psi: G\times(\SoXSt)\rightarrow\SoXSt,\quad 
((u_1,u_2),(\rho,\theta,\phi))\mapsto (\rho+u_1,\theta,\phi+u_2)
\] 
for $(u_1,u_2)\in\U\times\U$ and $(\rho,\theta,\phi)\in\SoXSt$. In
writing this, we always assume the standard identification of the
groups $\U$ and $\So$.  Hence a basis of generators of the action are
the coordinate fields $\partial_\rho$ and $\partial_\phi$. These are
globally smooth vector fields on $\SoXSt$ and one can check that this
action is global, smooth and effective. The action degenerates at
those points where the vector field $\partial_\phi$ has a zero, namely
at the poles of the $\St$-factor given by $\theta=0,\pi$.  In the case
of $\Sth$, we have the following action
\[\Psi: G\times\Sth\rightarrow\Sth,\quad 
((u_1,u_2),(\chi,\lambda_1,\lambda_2))\mapsto
(\chi,\lambda_1+u_1,\lambda_2+u_2).\] The generators of the group are
the coordinate fields $\partial_{\lambda_1}$ and
$\partial_{\lambda_2}$. Here, as well, these are globally smooth
vector fields on $\Sth$. Indeed, the action is global, smooth and
effective. The action degenerates where either $\partial_{\lambda_1}$
or $\partial_{\lambda_2}$ have a zero, which is the case at
$\chi=0,\pi$. Now, it is a fact, quoted in \cite{chrusciel1990}, that
all other smooth effective global actions of $\U\times\U$ on any of
these manifolds must be equivalent, i.e.\ can only differ by an
automorphism of the group or by a diffeomorphism of the manifold to
itself.  Hence it is sufficient to have a single representation of the
action.

We will now formulate the actions above in an equivalent, but more
geometrical manner.  We note that the fields $W_1$, $W_2$, $W_3$
defined in \Eqsref{eq:KS_KillingBasis}, together with
\begin{equation*}
  \xi_3:=\partial_\rho,
\end{equation*}
form a basis of the Killing algebra in the spatially homogeneous case
on \SoXSt. Moreover, the action of the group $\U\times\U$ is generated
by $\{W_3, \xi_3\}$, i.e.\ the translation vector field along the
$\So$-factor of the manifold and a rotation of the \St-factor of the
manifold corresponding to an element of the Lie algebra of $\SO$.  In
the \Sth-case, let us introduce the vector fields
\begin{subequations}
  \label{eq:S3referencefields}
  \begin{eqnarray}
    Y_1&=&2\sin \rho_1\,\partial_\chi
    +2\cos \rho_1\,
    \left(\cot\chi\,\partial_{\rho_1}-\csc\chi\,\partial_{\rho_2}\right),\\
    Y_2&=&2\cos \rho_1\,\partial_\chi
    -2\sin \rho_1\,
    \left(\cot\chi\,\partial_{\rho_1}-\csc\chi\,\partial_{\rho_2}\right),\\
    Y_3&=&2\partial_{\rho_1},\\
    Z_1&=&-2\sin \rho_2\,\partial_\chi
    -2\cos \rho_2\,
    \left(\cot\chi\,\partial_{\rho_1}-\csc\chi\,\partial_{\rho_2}\right),\\
    Z_2&=&2\cos \rho_2\,\partial_\chi
    -2\sin \rho_2\,
    \left(\cot\chi\,\partial_{\rho_1}-\csc\chi\,\partial_{\rho_2}\right),\\
    Z_3&=&2\partial_{\rho_2},
  \end{eqnarray}
\end{subequations}
in terms of the Euler angle coordinates above, but with $\rho_1$ and
$\rho_2$ defined by
\begin{equation}
  \label{eq:defrho}
  \lambda_1=(\rho_1+\rho_2)/2,\quad\lambda_2=(\rho_1-\rho_2)/2.
\end{equation}
The definitions of these fields are in agreement with the conventions
in \cite{beyer08:code,beyer08:TaubNUT}. These vector fields have the
property that they are globally smooth on \Sth and are invariant under
the standard left- and right-actions, respectively, of the group
$\text{SU}(2)$ on \Sth. Furthermore, they satisfy
\[[Y_a,Y_b]=2\sum_{c=1}^3\eta_{abc}Y_c,\quad
[Z_a,Z_b]=2\sum_{c=1}^3\eta_{abc}Z_c,\quad [Y_a,Z_b]=0.\] Here
$\eta_{abc}$ is the totally antisymmetric symbol with
$\eta_{123}=1$. Note that the collections $\{Y_1,Y_2,Y_3\}$ and
$\{Z_1,Z_2,Z_3\}$ are global frames on \Sth.  A basis of the
generators of the $\U\times\U$-action on $\Sth$ is $\{Y_3,Z_3\}$. We
remark that both $Y_3$ and $Z_3$ are nowhere vanishing vector
fields. They become, however, collinear at $\chi=0,\pi$, and the
$\U\times\U$-action degenerates.

Now, recall that both \SoXSt and \Sth are principal fiber bundles over
\St with structure group $\U$. In the \SoXSt-case, it is the trivial
bundle with bundle map
\[\Phi_1:\SoXSt\rightarrow\St,\quad (p,q)\mapsto q.\]
In the \Sth-case, it is the Hopf bundle and the bundle map can be
given the representation
\begin{align*}
  \Phi_2&:\Sth\rightarrow\St,\quad
  (x_1,x_2,x_3,x_4)\mapsto 
  (y_1,y_2,y_3)\\
  &=(-2(-x_1 x_3 + x_2 x_4), 
  2 (x_2 x_3 + x_1 x_4), x_1^2 + x_2^2 - x_3^2 - x_4^2).
\end{align*}
Here the notation is such that $(x_1,x_2,x_3,x_4)\in\Sth\subset\R^4$
with standard Cartesian coordinates on $\R^4$, and
$(y_1,y_2,y_3)\in\St\subset\R^3$. When we write this map in terms of
the Euler angle coordinates for \Sth, given by
\Eqsref{eq:eulerangleparm} and \eqref{eq:defrho}, and polar
coordinates on $\St$ given by \Eqref{eq:polarcoordinatesS2} (with
$x_i$ substituted by $y_i$), then the Hopf map has the 
simple representation
\[\Phi_2:(\chi,\rho_1,\rho_2)\mapsto (\theta,\phi)=(\chi,\rho_1).\]
In order to simplify the notation we write now
\begin{enumerate}
\item $N$ for $\SoXSt$ or $\Sth$,
\item $\eta$ for $\xi_3$ or $Z_3$, and $\sigma$ for $2W_3$ or $Y_3$,
\item $\Phi$ for $\Phi_1$ or $\Phi_2$,
\end{enumerate}
in the following. Furthermore, we write $G=\U\times\U$, and $G_1=\U$
to denote the subgroup of $G$ generated by $\eta$. Recall that $\eta$
is nowhere vanishing and note that its integral curves are closed
circles. We notice that in both cases, the field $\eta$ is tangent to
the fibers of the bundles, and hence $G_1$ becomes the structure
group, so that $N/G_1\cong\St$. Note that, in particular,
$\Phi_*\sigma$ is a smooth global vector field on $\St$, and indeed
equals the coordinate vector field $2\partial_\phi$. This fact holds
irrespective of the choice of either \SoXSt or \Sth, and thus clearly
demonstrates how similar the two cases are at this level.

The bundle maps $\Phi$ can be used as follows. Assume that we want to
solve a set of partial differential equations on $N$, possibly with an
additional time function $t$. We suppose that all unknowns and
coefficients are smooth functions on $N$ constant along $\eta$, and
that all differential operators in the equations originate in smooth
vector fields on $N$ whose Lie brackets with $\eta$ vanish. Then it
turns out that there is an equivalent system of equations on \St as
follows. Each such unknown and coefficient can be identified uniquely
with a smooth function on \St by means of the bundle map
$\Phi$. Furthermore, each such vector field yields a unique smooth
vector field on \St (which certainly has zeroes) when pushed forward
with $\Phi$. Most importantly, the solution of these new equations on
\St, if it exists, hence yields a unique corresponding solution of the
original problem on $N$. Thus without loss of information, we are
allowed to ``transport such geometric problems from $N$ to \St'' along
the bundle map $\Phi$.